\documentclass[useAMS,usenatbib]{mn2e}
\usepackage{lscape}
\usepackage{amsmath}
\usepackage{latexsym}
\usepackage{graphicx}
\usepackage{amssymb}
\usepackage{float}

\newcommand{\apj}{ApJ}
\newcommand{\mnras}{MNRAS}
\newcommand{\apjl}{ApJL}
\newcommand{\aj}{AJ}
\newcommand{\aap}{A\&A}
\newcommand{\apjs}{ApJS}
\newcommand{\lya}{Ly$\alpha$}
\newcommand{\zz}{$z\sim$ 2}
\newcommand{\zzz}{$z\sim$ 3}
\newcommand{\zzzz}{$z\sim$ 4}

\newcommand{\fesc}{$f_{\rm esc}^{\rm LyC}$}

\newcommand{\fescfif}{$f_{\rm esc}^{\rm 1500}$}
\newcommand{\fescrel}{$f_{\rm esc,rel}$}
\newcommand{\fobs}{$(F_{\rm UV}/F_{\rm LyC})^{\rm obs}$}
\newcommand{\robs}{$R_{obs}$}
\newcommand{\robsl}{$R_{obs}(\lambda)$}
\newcommand{\robsu}{$R_{obs}(U_n)$}
\title[Lyman continuum galaxies] {Lyman continuum galaxies and the
escape fraction of Lyman break galaxies}

\author[Cooke et al.]{J. Cooke$^{1}$\thanks{E-mail:
jcooke@astro.swin.edu.au}, E. V. Ryan-Weber$^1$, T. Garel$^{1,2}$,
and C. G. D\'{i}az$^1$\\
$^1$Centre for Astrophysics and Supercomputing, Swinburne University
of Technology, Hawthorn, VIC, 3122, Australia\\
$^2$Australian Research Council Super Science Fellow}

\begin{document}

\date{Accepted 2014 March 31. Received 2014 March 13; in original
form 2013 December 20}
\pagerange{\pageref{firstpage}--\pageref{lastpage}} \pubyear{0000}

\maketitle

\label{firstpage}

\begin{abstract} 

Lyman break galaxies (LBGs) at $z\sim$ 3--4 are targeted to measure
the fraction of Lyman continuum (LyC) flux that escapes from high
redshift galaxies.  However, $z\sim$ 3--4 LBGs are identified using
the Lyman break technique which preferentially selects galaxies with
little or no LyC.  We re-examine the standard LBG selection criteria
by performing spectrophotometry on composite spectra constructed from
794 $U_nG\cal{R}$-selected \zzz\ LBGs from the literature while adding
LyC flux of varying strengths.  The modified composite spectra
accurately predict the range of redshifts, properties, and LyC flux of
LBGs in the literature that have spectroscopic LyC measurements while
predicting the existence of a significant fraction of galaxies outside
the standard selection region.  These galaxies, termed Lyman continuum
galaxies (LCGs), are expected to have high levels of LyC flux and are
estimated to have a number density ${\sim30}$--50 percent that of the
LBG population.  We define \robsu\ as the relative fraction of
observed LyC flux, integrated from 912\AA\ to the shortest restframe
wavelength probed by the $U_n$ filter, to the observed non-ionising
flux (here measured at 1500\AA).  We use the 794 spectra as a
statistical sample for the full \zzz\ LBG population, and find
${\mbox{\robsu}=5.0^{+1.0}_{-0.4} ~(4.1^{+0.5}_{-0.3})}$ percent,
which corresponds to an intrinsic LyC escape fraction of
${\mbox{\fesc}=10.5^{+2.0}_{-0.8}~(8.6^{+1.0}_{-0.6})}$ percent
(contamination corrected).  From the composite spectral distributions
we estimate ${\mbox{\robsu}\sim16\pm3,~\mbox{\fesc}\sim33\pm7}$
percent for LCGs and
${\mbox{\robsu}\sim8\pm3,~\mbox{\fesc}\sim16\pm4}$ percent for the
combined LBG + LCG \zzz\ sample.  All values are measured in apertures
defined by the UV continuum and do not include extended and/or offset
LyC flux.  A complete galaxy census and the total emergent LyC flux at
high redshift is essential to quantify the contribution of galaxies to
the ionising photon budget of the Universe, particularly during the
epoch of reionisation.

\end{abstract}



\section{Introduction}\label{intro}

Since at least $z\sim$ 6, galaxies have been surrounded by ionised
hydrogen \citep{fan06,komatsu11,zahn12}, maintained by radiation from
quasars and galaxies.  The ultraviolet (UV) background changes over
time as the two populations evolve
\citep[e.g.][]{becker07,becker13,calverley11,mcquinn11,wyithe11},
however their relative contribution to the ionising spectrum is not
well constrained \citep{haardt12}.  As predicted from the declining
space density of quasars with redshift
\citep{hopkins07,jiang08,fontanot12}, hard radiation from quasars at
$z\lesssim$ 3 dominates the cosmic hydrogen photoionisation, whereas
beyond $z\sim3$ star-forming galaxies provide a softer, dominant
background.  Lyman-continuum (LyC; $<912$\AA) photons, capable of
ionising neutral hydrogen, are readily absorbed by hydrogen, helium,
and dust in the interstellar medium (ISM) of the producing galaxy.
Nevertheless, a fraction of these ionising photons survive their
journey through the ISM and contribute to the intergalactic UV
background.

The fraction of LyC photons escaping galaxies, \fesc, is fundamentally
tied to many astrophysical processes involved in the formation and
evolution of galaxies, including the reionisation of intergalactic
hydrogen and the suppression of gas collapse and star formation in
low-mass galaxies
\citep{efstathiou92,barkana99,bullock00,somerville02}.  Theoretical
calculations find that the value of \fesc\ has a high intrinsic
scatter \citep{fernandez11}, can range from 1 to 100 percent
\citep[e.g.][]{ricotti00,wood00}, and is higher on average in dwarf
galaxies \citep{yajima11} due to a more porous H\textsc{i} gas
distribution \citep[although see][]{gnedin08}.  The expected trend of
\fesc\ with redshift is similarly model-dependent; \citet{ricotti00}
state that \fesc\ of a fixed halo mass decreases with increasing
redshift, whereas \citet{razoumov06,razoumov10} find that \fesc\
increases from 1, to 10, to 80 percent at $z\sim$ 2, 4 and $10$,
respectively.

Analytic models and cosmological simulations often adopt \fesc\ =
20--50 percent for galaxies during the epoch of reionisation
\citep{iliev07,madau99,pawlik09,shull12}.  Contemporary simulations of
the epoch of reionisation find that the functional form of \fesc\ must
increase steeply with redshift to match current constraints on
reionisation and to maintain an ionised IGM
\citep{finlator12,kuhlen12}.  In order to reconcile stars as the main
contributor to ionising radiation during this epoch, the galaxy
luminosity function at $z\sim$ 7--9 must be extrapolated to absolute
magnitudes of M$_{UV}=-13$ or fainter (4 magnitudes fainter than
current observational limits at these redshifts) and an \fesc\ of
20--50 percent adopted \citep{robertson13,trenti10,finkelstein10}.

Despite stars being the leading candidate for the source of
reionisation, tension exists with the metallicity of the IGM at
$z\sim6$, which provides an independent estimate of the past total sum
of ionising photons from star formation.  The value of \fesc\ must be
$\gtrsim$50 percent to reconcile the measurement of the IGM
metallicity with the current census of ionising photons from galaxies
at $z\sim$ 6 \citep{ryan-weber09}.  Thus, efforts for a complete
census of high redshift star-forming galaxies and constraints on
\fesc\ are key to understanding how radiative feedback influences the
thermal and ionization state of the intergalactic medium and its
effect on galaxy formation and the reionisation of the Universe.

Measuring \fesc\ is not an easy task as the intrinsic LyC emission
from a galaxy is always attenuated.  \citet{dove94} analyze OB
associations in the Galactic plane and conclude that 7 percent of the
LyC flux escapes normal to the disk of the Milky Way into the halo.
Similar results, \fesc\ = 1--2 percent, when averaged over solid
angle, are obtained by measuring the H$\alpha$ emission of
photoionised high velocity clouds \citep{blandhawthorn99,putman03}.
For other galaxies, and galaxies at high redshift, the intrinsic value
of \fesc\ is impossible to measure and can only be estimated from the
surviving LyC and non-ionising flux reaching Earth and the estimated
amount of absorption by intervening gas in the line of sight.

Due to the very nature of the LyC being in the far-UV, ground-based
observational efforts for direct detection have focused on the
redshift ``sweet-spot'' of $z\sim$ 3--4 where the LyC is shifted to
the optical and yet a low enough redshift that the average number of
intervening absorption systems is not too great to hinder the
detection of LyC emission.

Galaxies at these redshifts are efficiently selected using the
Lyman-break technique
\citep{guhathakurta90,songaila90,steidel92,steidel93} which uses
broadband imaging to search for the predicted strong break in flux at
the Lyman limit (912\AA), when compared to UV continuum flux longward
of \lya\ (1216\AA), as a result of absorption in stellar atmospheres,
by the ISM of the galaxy, and by optically thick systems in the line
of sight.  Over the last two decades, follow-up spectroscopy has shown
that this technique is highly successful in selecting $z\sim$ 3--4
Lyman break galaxies (LBGs) and, when accounting for the statistical
change in \lya\ forest absorption with redshift, highly successful in
selecting $1\lesssim z\lesssim10$ LBGs while eliminating most low
redshift sources
\citep[e.g.,][]{steidel96,steidel99,steidel03,steidel04,cooke05,cooke06,bouwens07,bouwens11,ly11,bielby11,bielby13,ellis13}.
In this work, we refer to the \zzz\ and \zzzz\ LBG colour-selection
criteria of \citet{steidel03} and \citet{steidel99}, and all criteria
based on similar $<$ 912\AA\ flux expectations, as `standard'
criteria.

It is important to remind the reader that the standard LBG
colour-selection criteria were initially designed for {\it efficient}
galaxy detection at high redshift and not necessarily for {\it
comprehensive} detection.  Firstly, the standard criteria are
sensitive to star-forming galaxies that are luminous in the restframe
UV (initially to accommodate the higher sensitivities and
field-of-views of optical CCDs versus other wavelengths) and therefore
exclude many of the high redshift passive and dust-obscured
star-forming galaxies \citep[e.g.,][]{daddi04,chapman05,vandokkum06}.
Secondly, and more important to the work presented here, the standard
LBG colour-selection criteria are designed to select galaxies based on
the expectation of essentially zero detectable flux blueward of
912\AA.  Therefore, it is understandable but interesting to note that
all spectroscopic searches for LyC flux in high redshift galaxies have
been performed on LBGs.  As we show below, galaxies with detectable
LyC flux can reside within the standard LBG colour-selection region
but we need to look more broadly to properly measure the LyC flux of
the full high redshift galaxy population.

Here, we present the colours of $z\sim$ 3--4 galaxies, determined from
spectrophotometry performed on composite LBG spectra with various
levels of flux added blueward of 912\AA.  We find that their colours
are highly sensitive to the amount and wavelength extent of the LyC
flux and place the galaxies inside and outside the standard LBG
colour-selection region.  The existence of galaxies at $z\sim$ 3--4
with colours outside the standard LBG colour-selection criteria have
been previously confirmed.  The VLT Vimos Deep Survey
\citep[VVDS][]{lefevre05a} is a magnitude-limited search for galaxies
down to $i'=$ 24.75, independent of colour selection constraints, that
has spectroscopically confirmed $\sim$400 $z\sim$ 3--4 galaxies inside
and outside the conventional LBG colour-selection criteria with
potentially similar space densities
\citep{lefevre05b,lefevre13a,lefevre13b,paltani07}.  We critically
review the VVDS data and report the results in a companion paper.

This paper presents the colour evolution of $z=$ 2.7--4.5 galaxies as
determined from the modified composite spectra that accurately predict
the redshifts, properties, and observed surviving LyC flux of LBGs
previously measured in the literature.  In addition, this work
provides a physical basis for the locations of spectroscopically
confirmed galaxies both inside and outside the standard LBG
colour-selection criteria with respect to their redshifts and \lya\
equivalent widths.  The predictive power of the modified composite
spectra has utility in estimating the global observed LyC flux from
LBG photometric samples and from LBG spectroscopic samples that do not
have deep LyC sensitivity.  Finally, the composite spectral analysis
predicts, and the data distribution and the VVDS confirmed spectra
demonstrate, that $z\sim$ 3--4 galaxies exist outside the standard
criteria and that the missed galaxies have moderate to strong
surviving LyC flux.

This paper is organised as follows.  \S2 discusses the observations
used in this work and \S3 describes the escape fraction and relevant
definitions.  We examine the LyC flux expectations of the data and of
the standard LBG colour-selection criteria in \S4.  We explore the
predictions of the modified composite spectra in \S5 and compare the
predictions to previous LyC flux searches and measure the average
observed LyC flux and \fesc\ of $z\sim$ 3 LBGs.  In \S6, we describe
the expectation for galaxies with colours outside the standard LBG
criteria and estimate their observed LyC flux and \fesc\ and that for
the full $z\sim$ 3 population.  We summarise the results in \S7.  All
magnitudes are reported in the AB magnitude system \citep{fukugita96}.


\section{Observations}\label{obs}

\begin{figure}
\begin{center}
\scalebox{0.40}[0.40]{\rotatebox{90}{\includegraphics{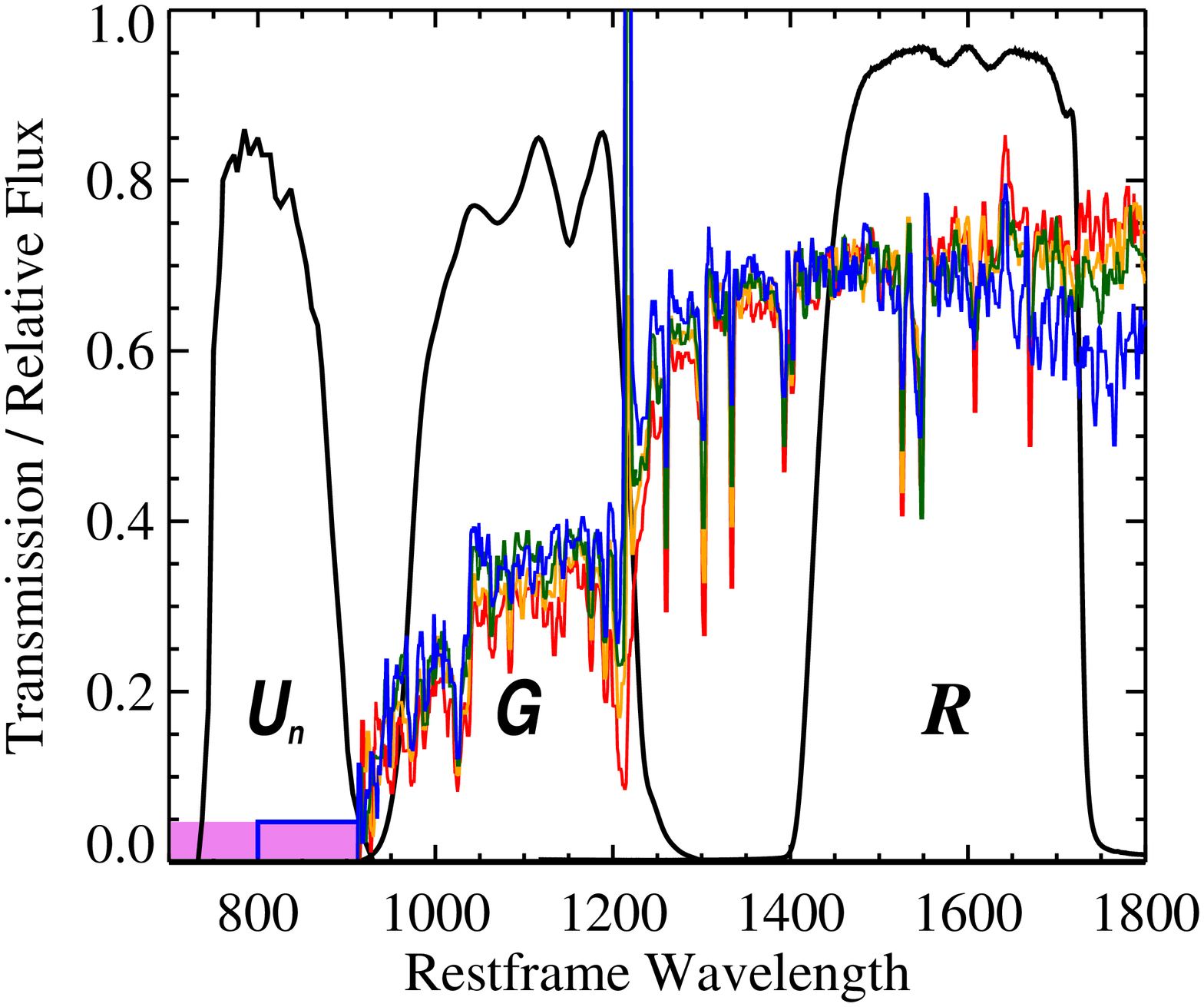}}}

\caption {\small Composite \zzz\ LBG spectra and transmission curves
for the $U_nG\cal{R}$ filters of \citet[][black curves]{steidel03}.
The composite spectra consist of $\sim$200 LBGs based on their \lya\
EW \citep{shapley03} and are shown arbitrarily normalised to their
flux at 1500\AA.  The filter transmission curves reflect the restframe
wavelength coverage for objects at ${z=3.3}$.  Above ${z\sim3.3}$, the
$U_n$ filter probes Lyman continuum flux exclusively.  Two models for
the coarse form of the escaping Lyman continuum flux as detected by
the broadband filters are shown (with nonspecific strengths): (1) a
`step' model (blue lines) that extends down to 800\AA\ and (2) a
`flat' model (violet region) that extends beyond the range of the
$U_n$ bandpass. }

\label{UnGR}
\end{center}
\end{figure}

To place this work on common ground, we perform our analysis on the
\zzz\ LBG $U_nG\cal{R}$ photometric data
set\footnote{http://cdsarc.u-strasbg.fr/viz-bin/Cat?J/ApJ/592/728} of
\citet[][hereafter S03]{steidel03} and use four composite spectra
constructed from the LBG quartiles of \citet{shapley03}.  The
composite spectra consist of 198 (or 199) LBG spectra from the S03
sample and are categorised by their net \lya\ equivalent width (EW).
Strong relationships have been found between \lya\ EW and several
other LBG properties such as ISM absorption line strength/velocity, UV
continuum slope, UV and optical morphology, and both small- and
large-scale environment \citep{shapley03,law07,law12,cooke10,cooke13}.
Thus, the \lya\ feature is a strong tracer of multiple observed
restframe UV and optical properties and environments of LBGs.  The
transmission of the $U_nG\cal{R}$ filters and the four composite
spectra are shown in Figure~\ref{UnGR}.

We have computed the magnitudes and colours of our modified composite
spectra in the $U_nG\cal{R}$ filter bandpasses because they are
constructed from the S03 data set.  However, the results of this work
are not unique to the $U_nG\cal{R}$ filters.  The composite spectra
can be ``observed'' in other filters when convolving the flux with the
relevant filter transmission, CCD quantum efficiency, and atmospheric
extinction.  We perform such an analysis using the
Canada-France-Hawaii Telescope Legacy Survey MegaCam $u^*g'r'i'z'$
filter set in \S\ref{contam}.


\section{The escape fraction}\label{exp}

The absolute escape fraction of LyC photons cannot be measured
directly as a result of the unknown absorption by intervening neutral
gas in the line of sight and the unknown true number of ionising and
UV continuum photons produced by a given galaxy.  Instead, previous
studies have focused on measuring the relative fraction of
non-ionising UV continuum photons to escaping LyC photons, \fescrel,
when combined with empirical and/or simulated estimates of the
intrinsic galaxy spectral energy distribution and absorption of the
IGM \citep[e.g.][]{steidel01,inoue05,shapley06,iwata09}.  The
parameter \fescrel\ is conventionally defined as
\begin{equation}\label{fesc-def} f_{\rm esc,rel}=\frac{f_{\rm
esc}^{LyC}}{f_{\rm esc}^{UV}}=\frac{(L_{\rm UV}/L_{\rm LyC})^{\rm
int}}{(F_{\rm UV}/F_{\rm LyC})^{\rm obs}} ~\rm{exp}(\tau^{\rm
IGM}_{\rm LyC}) \end{equation}

\noindent where \fesc\ and $f_{\rm esc}^{UV}$ are the intrinsic escape
fractions of photons in the Lyman continuum and UV wavebands,
respectively; $(L_{\rm UV}/L_{\rm LyC})^{\rm int}$ is the fraction of
intrinsic non-ionising to ionising luminosity density produced by the
galaxy; \fobs\ is the observed non-ionising to ionising flux density
fraction; and $\rm{exp}(\tau^{\rm IGM}_{LyC})$ is the factor
describing the attenuation of the LyC flux by the IGM.

Any interpretation of the observed flux ratio, \fobs, as a proxy for
the intrinsic $f_{\rm esc}^{\rm LyC}$ requires assumptions or modeling
of $(L_{\rm UV}/L_{\rm LyC})^{\rm int}$, $\rm{exp}(\tau^{\rm IGM}_{\rm
LyC})$ and $f_{\rm esc}^{\rm UV}$.  For example, \citet{inoue05}
estimate $(L_{1500}/L_{900})^{\rm int}=$ 1--5.5 for a wide range of
stellar populations, with a typical value in the literature set at
3.0.  The attenuation from intervening optically thick systems has
been estimated analytically \citep[e.g.,][]{inoue05}, via Monte Carlo
simulations \citep[e.g.,][]{shapley06}, and empirically using QSO
spectra \citep[e.g.,][]{steidel01,prochaska09} with consistent
results.  Finally, $f_{\rm esc}^{\rm UV}$ is typically computed by
applying a dust law to the measured, or estimated, $E(B-V)$ of the UV
continuum.  From equation~\ref{fesc-def} we can see that, if we assume
$(L_{\rm UV}/L_{\rm LyC})^{\rm int}$, ${\rm exp}(\tau^{\rm IGM}_{\rm
LyC})$ and $f_{\rm esc}^{UV}$ remain more or less constant, a smaller
\fobs\ implies a larger \fesc.

The only quantity in equation~\ref{fesc-def} that can be observed for
individual galaxies (or measured in the composite spectra) is \fobs.
With the considerations above, we see that the Lyman break feature of
LBGs favours small observed LyC flux corresponding to large \fobs.  As
a result, LBGs are non-ideal galaxies to search for high levels of
escaping LyC flux.  As we show below, only LBGs having unexpected
colours relative to their spectral type and redshifts are expected to
show measurable, or high, levels of escaping LyC photons.

Two spectroscopic measurements of the mean surviving LyC flux of
$U_nG\cal{R}$-selected \zzz\ LBGs were performed by \citet{steidel01}
and \citet{shapley06}.  In both cases, the results are reported using
\fobs = $F_{1500}/F_{900}$, where $F_{1500}$ is the flux density
measured at 1500\AA\ and $F_{900}$ is the flux density averaged over
880--910\AA.  \citet{steidel01} find ${\mbox{\fobs}=17.7\pm3.8}$ for a
sample of 29 LBGs at ${\langle z\rangle=3.40\pm0.09}$, or an \fescrel\
$\gtrsim$ 50 percent when using their empirical IGM absorption
estimate and adopted $(L_{\rm UV}/L_{\rm LyC})^{\rm int}=$ 3.0.  The
\citet{steidel01} sample has been thought to be biased in that the
LBGs are pulled from the bluest quartile (quartile 4).  Indeed as we
show below, the locations in colour-colour space of these LBGs are
offset with respect to the `standard' expectations for $z=3.4$ LBGs
from star-forming templates and, in particular, offset with respect to
the the expectations of $z=3.4$ quartile 4 LBGs.

\citet{shapley06} investigate a sample of fourteen $\langle
z\rangle=3.06\pm0.12$ LBGs more representative of the colour and \lya\
EW of \zzz\ LBGs and find an average $F_{1500}/F_{900}=
58\pm18_{stat}\pm17_{sys}$ corresponding to an \fescrel\ $\sim$ 13
percent.  These results include two exceptional cases,
$F_{1500}/F_{900}\mbox{(C49)}= 12.7\pm1.8$ (\fescrel\ = 65 percent)
and $F_{1500}/F_{900}\mbox{(D3)}= 7.5\pm1.0$ (\fescrel\ $\sim$ 100
percent), that contribute nearly all of the LyC flux to the average.
Subsequent observations, however, reveal that in one case the observed
LyC flux may be line-of-sight contamination and in the other case the
result of scattered light\footnote{The apparent LyC flux from C49 has
been found to arise from a lower redshift object contaminating the
field \citep{nestor13}.  Thus it is likely that this object has little
to no escaping LyC flux.  \citet{nestor13} detect no LyC flux from
object D3 to the limits of their data.  The source of the observed LyC
flux in the deep spectroscopy is still unresolved but may be caused by
a nearby bright object as discussed in \citet{shapley06}.}.  Omitting
these two galaxies from the sample places $F_{1500}/F_{900}\sim$ 400
and \fescrel\ $\sim$ 2 percent.  The LBGs in this sample have colours
that are expected for their redshifts and spectral properties,
i.e. colours that are expected for galaxies having a strong Lyman
break, and thus have low expected LyC escape fractions.

Efforts using other selection criteria and filters, including
narrowband imaging, to measure the \zzz\ escape fraction have reported
results that span a large range.  \citet{iwata09} measure \fobs\ =
2.4--23.8 for seven LBGs from narrowband imaging.  The median, \fobs\
= 6.6, corresponds to \fescrel\ = 16--83 percent under various
assumptions detailed in that work.  \citet{inoue05} place upper limits
of \fescrel\ $<$ 72 percent and \fescrel\ $<$ 216 percent for two
LBGs.  Recently, \citet{nestor13} combined narrowband imaging and
spectroscopic samples of LBGs and \lya\ emitting galaxies (LAEs) and
find model dependent \fescrel\ $\sim$ 25--35 percent and \fescrel\
$\sim$ 33--100 percent for the two galaxy types, respectively, taking
into account the typical ages and metallicities of these galaxies.

To help alleviate potential confusion between various observational
and theoretical definitions of the escape fraction, and to emphasise
that the quantities we are measuring (or predicting) here are the LyC
and non-ionising UV continuum flux {\it observed}, we define the
relative fraction 
\begin{equation}\label{robs-def} 
R_{obs}(\lambda) \equiv \frac{F^{LyC}_{obs}}{F^{UV}_{obs}},
\end{equation}

\noindent where $F^{LyC}_{obs}$ is the observed LyC flux density
integrated from $\lambda$ to 912\AA\ and $F^{UV}_{obs}$ is the
observed non-ionising UV continuum flux density.  The predicted values
of \robsl\ in this work are based on $F^{LyC}_{obs}$ integrated over
the $U_n$-band, unless otherwise noted.  The wavelength is not
specified here because it will vary within the $U_n$ filter with the
redshift of the galaxies (or vary by model).  For the work presented
here, we compute $F^{UV}_{obs}$ from the composite spectra flux at
1500\AA.

From an observational point of view, expressing the surviving LyC flux
as \robsl\ is more intuitive and, because the observed UV continuum is
measured more accurately, the error in \robsl\ goes as the uncertainty
of the observed LyC flux.  We note that \robsl\ is not to be confused
with \fesc, although both are expressed as percentages.  Below, we
present the colours of $z\sim$ 3--4 galaxies as a function of redshift
and spectral type and compare the results to the expectations of the
standard LBG colour selection criteria in an effort to shed light on
the results of escaping LyC flux measurements in the literature.


\section{Conventional expectations}\label{conv_exp}

The standard LBG colour-selection criteria is based on the colour
evolution of star-forming galaxy templates as a function of redshift.
Lines are drawn in colour-colour space in an effort to mark off areas
that include the galaxy template predictions (including photometric
uncertainty estimates) at the desired redshifts while excluding low
redshift sources.  Because the colours of high redshift and low
redshift galaxies overlap in colour-colour space, border regions
become a compromise between `efficient' criteria, that minimize the
fraction of low redshift sources at the cost of excluding a fraction
of the high redshift population, and `comprehensive' criteria that aim
to include all high redshift galaxies over a specified redshift range
at the cost of including a significant fraction of low redshift
sources.

To assess the expectations of the standard criteria {\it a
posteriori}, we compute the spectrophotometric colours of the four
composite LBG spectra and analyze their colour-colour distribution
with that of the data.  Here, we assume no flux shortward of 912\AA\
for our composite spectra to match template-based, Lyman-break
criteria.  As stated in \S\ref{obs}, the composite spectra are
constructed from the LBG sample of S03 and are, thus, selected by
their $U_nG\cal{R}$ colours.  We generate four composite spectral
samples by randomly pulling values from the observed $\cal{R}$
magnitude and redshift distributions of the data that form each LBG
quartile and assign those values to the respective composite spectrum.
We then compute their flux in the $U_n$ and $G$ bandpasses and include
their associated photometric uncertainties.

To check the accuracy of the spectrophotometry, we examine the
distributions of the composite spectral samples on the $(G-\cal{R})$
vs.\ $\cal{R}$ colour magnitude diagram.  The means and dispersions of
the composite spectral sample distributions are compared to the data
for each respective quartile.  We find very close ($\le$ 0.02
magnitude) agreement when examining the distributions quartile by
quartile and as a full photometric sample.  The results can be found
in Table 1 of \citet{cooke13} along with a more detailed discussion.
The agreement confirms the accuracy of the spectrophotometry and
provides an assurance that, for this purpose, each of the four
composite spectra is accurate average representation of the 25 percent
of the LBG population in which they are comprised.  Only the composite
spectral sample with the strongest \lya\ emission (quartile 4) shows a
small departure from its respective spectroscopic sample, with its
colour distribution mean 0.12 mags brighter in the $G$-band where the
\lya\ feature falls.  This discrepancy is caused by a small number of
strong \lya\ emitting LBGs adding a disproportionate weight to the
average spectrum profile.  We apply a 0.12 magnitude decrease in
$G$-band flux to bring the composite spectrum in line with the average
values of that quartile.  We note that not correcting for this effect
results in a larger discrepancy in the distributions discussed below,
thus our analysis is conservative.

\begin{figure}
\begin{center}
\scalebox{0.30}[0.30]{\rotatebox{90}{\includegraphics{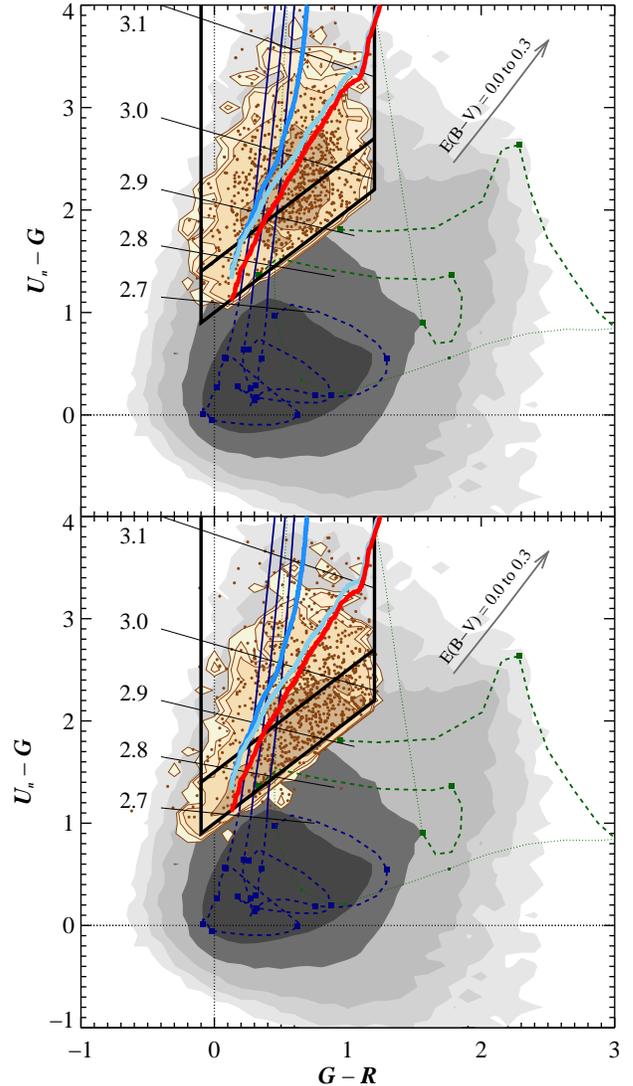}}}

\caption {\small Colour-colour plots of the \zzz\ Lyman break galaxies
(LBGs) of \citet[][S03]{steidel03}.  Grey contours reflect the general
colours of field objects with each contour 4$\times$ the density of
the next darker (interior) contour.  Dotted (green) curves denote the
colour evolution of two early type galaxy templates from $z=$ 0--2
($z>$ 1 colours are shown as thinner curves and are less reliable).
Dashed (dark blue) curves trace the colour evolution of star-forming
galaxy templates for $z=$ 0--2.7 and change to solid curves for $z>$
2.7.  Squares mark each template at intervals of $\Delta z=$ 0.5 and
angled lines mark the average redshifts of the star forming templates
at $z>$ 2.7 and are labelled.  The thick black lines enclose the S03
\zzz\ colour-selection criteria.  The thick coloured curves are the
evolutionary tracks for the four composite spectra (averaged) having
zero flux $<$ 912\AA\ using three prescriptions: (1) theoretically
infinite magnitude detections (mid-blue), (2) assignment of $U_n$-band
`non-detections' to the limiting $U_n$ magnitude of the S03
observations (light blue) and (3) the inclusion of
wavelength-dependent slit flux loss and a $U_n$-band red leak to
prescription 2 {\it (red; see text)}.  The arrows indicate the
direction and magnitude that reddening affects the colours.  The
tan/brown contours (relative 4$\times$ density levels) show the
expected LBG distribution based on prescription 3 {\it (upper panel)}
and the actual distribution of the data {\it (lower panel)}.  Filled
circles indicate the spectroscopically confirmed LBGs {\it (lower
panel)} and an equal number of random composite spectra values {\it
(upper panel)}. }

\label{conventional}
\end{center}
\end{figure}

Our next step is to examine the composite spectral sample and data
distributions on the $(U_n-G)$ vs.\ $(G-\cal{R})$ plane.  Although the
composite spectra are constructed directly from the S03 data, we find
that the distributions differ (Figure~\ref{conventional}).  The
composite spectral samples have a relatively smooth distribution
throughout the standard colour-selection region with a density that
peaks near the redshift distribution peak ($\langle z\rangle=$ 2.96,
1$\sigma=$ 0.29; S03) in accordance with the expectations of the
templates, whereas the data distribution is seen crowded to bluer
$(U_n-G)$ colours and has a peak density near the
${(U_n-G)=(G-\cal{R})\mbox{~+ 1}}$ colour selection border.  We note
that the composite spectral sample distribution in
Figure~\ref{conventional} is the distribution after corrections
(discussed below) which act to push the distribution to bluer
$(U_n-G)$ colours and closer to that of the data.  The uncorrected
distribution has larger $(U_n-G)$ colours and a larger discrepancy
with the data.  The arrows in the figure indicate the direction and
magnitude in which reddening acts on the distributions, which is
essentially perpendicular to the direction of the discrepancy.  Thus,
the difference in the two distributions lies in their $(U_n-G)$
colours and, more specifically, in the $U_n$ magnitudes of the
galaxies.

Factors that can lead to a difference in $U_n$ magnitudes include (1)
a $U_n$-band red leak, (2) assignment of $U_n$-band non-detections to
an arbitrary magnitude (i.e., the limiting magnitude of the survey),
(3) the effects of wavelength-dependent flux loss in the spectroscopy
in which the composite spectra are formed, and/or (4) the initial
condition of zero flux shortward of 912\AA.  We assess each effect in
turn, below.

\subsection{$U_n$-band red leak}

The $U_n$ filter used in the S03 survey suppresses essentially all
flux redward of 4000\AA\ (Steidel 2013, private communication).  Only
a negligible flux leak appears in the filter transmission curve
occurring near 11,000\AA, $\sim$700\AA\ FWHM, with a flux peak
$\sim$0.01 of full transmission.  The leak resides largely beyond the
sensitivity of the CCDs used in that work and would result in a red
leak contribution of $\lesssim$0.01 magnitude, even when considering
the most efficient CCD quantum efficiency at long wavelengths and
brightest LBGs.  Redleak is determined not to contribute to the
distribution discrepancy but we consider a 10$\times$ magnitude leak
(0.1 mag) below as a conservative measure and to test the overall
effect redleak has on the distribution.

\subsection{Limiting magnitude detections}

The finite depths of imaging surveys limit the range of measurable
colours.  The resulting photometry can only provide lower limits for
objects with large colours, such as those with little or no flux in
the $U_n$ band and appear as non-detections or `drop-outs'.  For the
S03 survey, $U_n$-band non-detections were assigned the 1$\sigma$
limiting $U_n$ magnitude of their respective field.  To match these
data, we similarly assign the $U_n$ limiting magnitude to the
generated composite spectra spectrophotometry for those having
computed $U_n$-band mags equal to, or fainter than, the field limits.

\subsection{Wavelength dependent spectroscopic flux loss}

The spectroscopic data were acquired using the Low-Resolution Imaging
Spectrometer \citep[LRIS;][]{oke95,steidel04} at the Keck Observatory
in multi-object spectroscopic mode using slitmasks prior to the
availability of an atmospheric dispersion corrector.  Because the
slitlets are fixed at a given position angle during integration, flux
can be lost due to atmospheric dispersion over the course of the
observation as the slitlet position angle changes with respect to the
parallactic angle.  Assessing potential wavelength-dependent flux loss
in the spectroscopic observations is difficult.  Fortunately,
considerable effort was made in the design and execution of the
observations to minimize this effect by designing the position angles
of the slitmasks near the parallactic angle at mid-observation,
observing the targets at low airmass and as they pass the meridian,
and limiting the length of the exposure times.  As a result, the
observations should have incurred little wavelength-dependent flux
loss.  Nevertheless, we estimate the flux loss and its impact on the
LBG colour-colour distribution.

The close, $\le$ 0.02 magnitude, agreement between the composite
spectral sample and photometric data distributions on the $G-\cal{R}$
vs.\ $\cal{R}$ colour-magnitude diagram demonstrates that the flux
loss between the $G$ and $\cal{R}$ bandpasses is small and provides a
constraint on the average amount of flux lost in the $U_n$ bandpass.
Given the amplitude of the flux loss constraint between the $G$ and
$\cal{R}$ bandpasses, we compute the range of flux loss in the $U_n$
bandpass from observations with slit position angles that change by
various amounts from 0--45 degrees off parallactic during 2 and 3 hour
long integrations while passing through 1.1--1.5 airmass corresponding
to the respective ranges in the observations (or greater).  We find
that 90 percent of the observations incur less than $\sim$20 percent
flux loss ($\lesssim$ 0.24 mag).  Here, we conservatively assume 0.24
magnitude flux loss for the entire sample.

\subsection{Excess $U_n$-band flux}

The results of the above analyses are shown in
Figure~\ref{conventional}.  The evolutionary track for the average of
the four composite spectra, uncorrected for any effect above, is shown
as the thick, near vertical, mid-blue curve.  This curve represents a
theoretical infinitely deep $U_n$-band survey in which all colours can
be measured.  The strengths of the effects on the $U_n$-band magnitude
discussed above on the composite spectral sample distributions is
traced in Figure~\ref{conventional} by the change in the mean
composite spectrum evolutionary track.

The thick light-blue curve is the track that includes the assignment
of the $U_n$-band non-detections to the limiting magnitude of the
fields.  This effect causes the largest change in the distribution.
The thick red curve incorporates both the $U_n$-band non-detection
assignments and a 0.34 magnitude increase to the $U_n$ flux of the
composite spectra to accommodate both slit flux loss and $U_n$-band
red leak (0.24 + 0.1 magnitude, respectively).  The contours in the
upper panel of Figure~\ref{conventional} outline the expected
distribution of S03 LBGs for the latter case: assigning $U_n$
magnitude limits, slit flux loss, and red leak contributions (i.e.,
the thick red track).  The distribution reveals that, although the
combination of these three effects make a noticeable difference, and
move the distribution in the direction of the data, their
contributions fall well short of resolving the distribution
discrepancy.  This leaves the likelihood that the bulk of the
difference between the two distributions is caused by a genuine excess
in $U_n$-band flux.

\subsubsection{Contamination by foreground line-of-sight sources}

This above result does not consider $U_n$-band flux contamination from
lower redshift sources in the line of sight.  We address this topic in
detail in \S\ref{contam} but mention a few points here.  The S03
photometry is measured in 2 arcsec apertures, which differs in nature
to narrow-band surveys that measure extended flux.  Any lower redshift
source would need to physically overlap the photometric aperture,
albeit only weakly (i.e., even the wings of a faint line-of-sight
source might offer a measurable contribution).  However, at least two
effects act to keep the contamination rate low, (1) $\cal{R}<$ 25.5
\zzz\ LBGs are small in size (faint point source-like objects) and
have a relative low surface density ($\sim$1.8 arcmin$^{-1}$) and (2)
some line-of-sight sources overlapping with LBG apertures contribute
to all colours such that they no longer meet the standard LBG
criteria.  As we show later, we find that the number of LBGs meeting
the standard criteria with measurable (i.e., down to a level of m
$\sim$ 28) LyC contamination is small and is insufficient to explain
the distribution mismatch.  As a result, the existence of unaccounted
$U_n$-band flux, combined with the fact that LBGs in the literature
have measured, non-negligible, observed LyC flux (\S\ref{intro}), make
it worthwhile to re-examine the LBG colour-selection criteria and the
colour expectations of galaxies with various levels of observed LyC
flux.


\section{Lyman continuum expectations}\label{analysis}

To determine the expectations for galaxies with surviving LyC flux, we
modify the composite spectra and a star-forming galaxy template by
adding flux using two models: (1) a `flat' model, that includes flux
from 912\AA\ to the shortest wavelength probed by the $U_n$-band,
depending on the galaxy's redshift, to explore the effects of flux
contributions extending to short wavelengths and (2) a `step' model
that includes flux between 800--912\AA, motivated by the observation
of apparent (or potential) flux down to $\sim$800\AA\ in
\citet{shapley06} and to test a range of wavelength between zero LyC
flux expectations (912\AA) and that of the 'flat' model.  The two
models are pictorially represented in Figure~\ref{UnGR}.  Because we
are measuring the integrated flux shortward of 912\AA\ using blunt
tools in the form of broadband filters, the details of erratic flux
profiles resulting from the multitude of IGM systems in the
line-of-sight are smoothed over.  As a result, \robsl\ reported here
reflects the fraction of LyC flux integrated over the relevant
wavelengths when compared to the flux measured at 1500\AA\ from the
composite spectra.  Thus, the `step' model is denoted as \robs(800\AA)
and the `flat' model is denoted as \robs$(U_n)$.

\begin{figure}
\begin{center}
\scalebox{0.3}[0.3]{\rotatebox{90}{\includegraphics{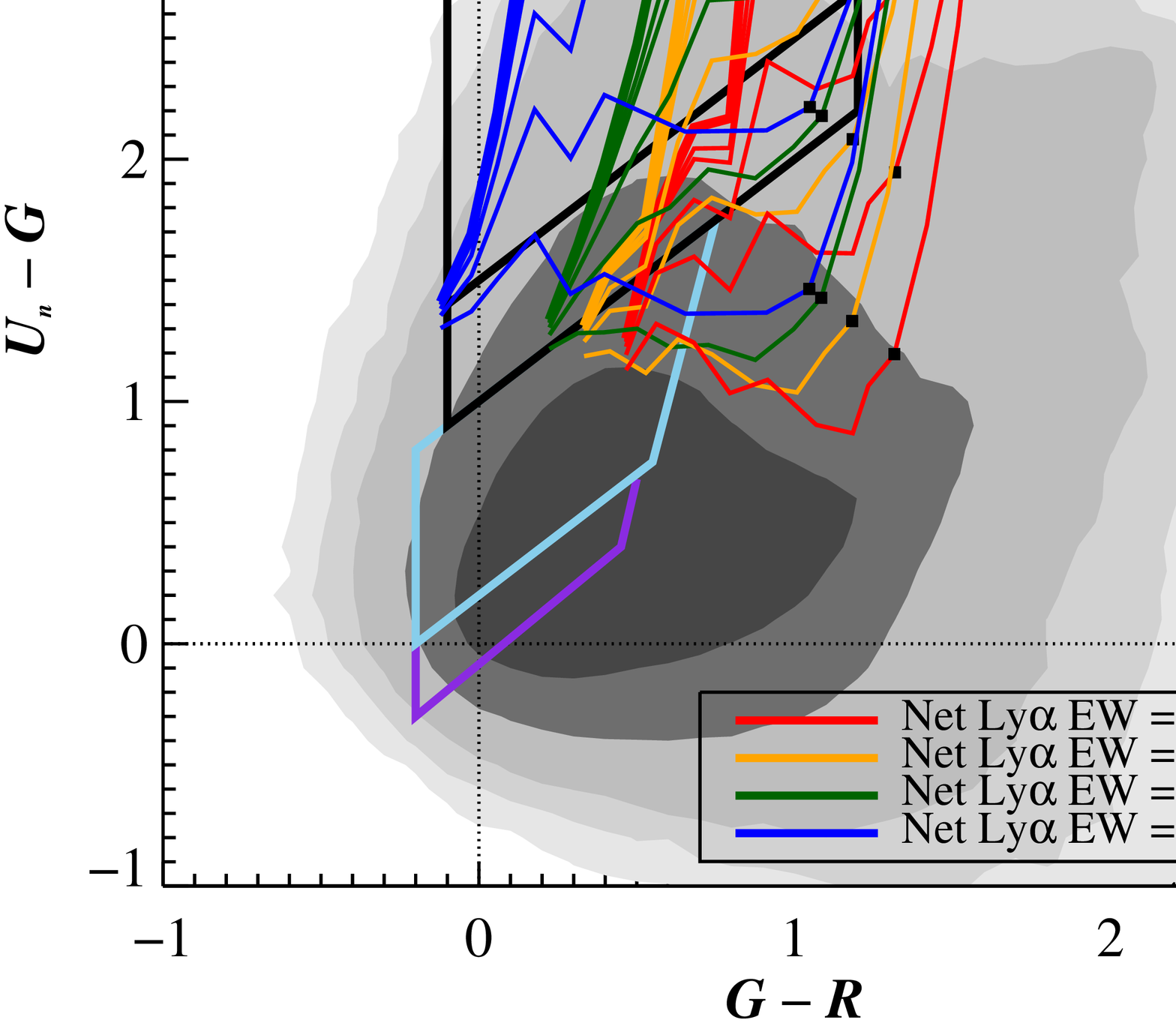}}}

\caption {\small Colour-colour plots similar to
Fig.~\ref{conventional}.  Here, the ${z\sim3}$ colour-selection region
is shown in each panel along with the ${z\sim2.2}$ (BX; light blue)
and $z\sim$ 1.7 (BM; purple) colour-selection regions
\citep{steidel04}.  In both panels, the evolutionary tracks of the
four composite spectra, quartile 1--4 of \citet{shapley03}, are shown
in red, orange, green, and blue, respectively, and redshifted from
${z=2.7-4.5}$ (the black squares denote $z$ = 3.5).  Composite spectra
with zero LyC flux are shown as the thick curves.  Peeling off from
these tracks (thin curves) downward and to the right in increasing
amounts are tracks for \robsl\ = 1, 2, 5, 10, and 20 percent (\fobs\ =
100, 50, 20, 10, and 5, respectively) as determined from two models:
{\it (upper panel)} a `flat' continuum extending from 912\AA\ down to
wavelengths below the $U_n$ bandpass and {\it (lower panel)} a step
function having flux from 800--912\AA.  The models show that {\it i)}
the galaxy colours are very sensitive to the integrated amount of
escaping LyC flux and {\it ii)} $z>$ 3.5 galaxies are very sensitive
to the wavelength extent of the escaping flux.  The models also show
that a significant fraction of galaxies are expected to reside outside
the standard colour-selection region.  Many of the `missed' galaxies
are outside lower redshift (BM and BX) selection criteria.  }

\label{z3}
\end{center}
\end{figure}

Figure~\ref{z3} shows the effects on the composite spectra
evolutionary tracks for the two models from $z=2.7$ to $z=4.5$ when
including various values of \robsl.  In both panels, the near
vertical, thick-lined evolutionary tracks represent zero flux below
912\AA\ colour-coded as red, orange, green, and blue corresponding to
quartiles 1--4 \citep{shapley03}, respectively.  The thin-lined tracks
represent composite spectra evolution with added LyC flux.  As \robsl\
is increased, the farther the tracks peel off from the zero LyC flux
tracks and down toward bluer $(U_n-G)$ colours.  It is readily
apparent that galaxy colours are very sensitive to even a small amount
of surviving LyC flux.

We extend the \zzz\ distributions used for the composite spectral
samples to $z=$ 3.5--4.5 (for reference, $z=$ 3.5 for each track is
marked with a black square in Fig.~\ref{z3}) by fading the $\cal{R}$
magnitude distribution according to redshift (e.g., a standard candle
is fainter by $\sim$0.5 magnitude from \zzz\ to \zzzz) and by
correcting the composite spectra for the change in flux decrement,
$D_A$, by \lya\ forest absorption with redshift.  For the latter, we
modify the \lya\ forest in the composite spectra using $D_A= (0.2
\cdot z - 0.6)$, derived from a fit to results in the literature
\citep{giallongo90,lu94,reichart01,jones12} and to return zero
correction for the $z=$ 3 composite spectra.

The difference between the two models lies in the shortest wavelength
that flux is assigned.  When considering $z=$ 2.7--4.5 galaxies, the
`step' model provides an approximate mid-point between the zero LyC
flux evolutionary tracks and the tracks of the `flat' model which span
the full range of outcomes.  Other observed LyC wavelength ranges can
be roughly interpolated between the expectations of the zero LyC flux
evolutionary tracks and those of the two models.

The $(U_n-G)$ colours of galaxies at all redshifts studied are
sensitive to both the amount and the wavelength extent of the LyC
flux, with the sensitivities increasing with redshift.  Inspection of
the `flat' model in the upper panel reveals that for $z>$ 3.5 galaxies
it only requires a small amount of LyC flux, \robs$(U_n)\lesssim$ 1
percent [\fobs\ $\gtrsim$ 100] to move the tracks from infinity (zero
LyC flux case) down to $(U_n-G)<$ 4 and only \robs$(U_n)\lesssim$ 5
percent [\fobs\ $\gtrsim$ 20] to move the tracks to $(U_n-G)<$ 2.  The
results are similar for the `step' model shown in the lower panel for
objects at $z\lesssim$ 3.8 where $>$ 800\AA\ restframe flux falls in
the $U_n$-band.

For both models, $z<$ 3.5 galaxies with small to moderate \robsl\ have
noticeable changes to their $(U_n-G)$ colour but remain in the
standard selection colour region.  However, higher amounts of LyC flux
can affect their $(U_n-G)$ colour such that a significant fraction of
galaxies fall out of the standard colour-selection region.  Many of
these `missed' galaxies evade lower redshift, e.g., BX and BM,
criteria and reside in a relatively unexplored region of high redshift
galaxy colour space.  This region is occupied by a higher density of
low redshift sources and presents complications in devising efficient
selection criteria.

The results for \zzzz\ galaxies in the $(G-\cal{R})$ vs. $(\cal{R}-I)$
colour-colour plane is very similar to that for \zzz\ shown here.
Typically, the $(U_n-G)$ vs.\ $(G-\cal{R})$ plane is not considered in
$z\sim$ 4 galaxy selection.  However, we find that isolating galaxies
on this colour-colour plane can be an effective approach to select
galaxies with measurable LyC flux at \zzzz\ from high-density regions
containing low redshift sources.  Galaxies at \zzzz\ are located in
the right half of Figure~\ref{z3} and beyond the $z=$ 3.5 markers
(small black squares).  Deep near-UV (NUV) imaging and a parallel
analysis on the $(NUV-U_n)$ vs.\ $(U_n-G)$ plane would enable \zz\ and
\zzz\ galaxy selection following a similar approach to that presented
here for \zzz\ and \zzzz.  Moreover, photometric redshifts using
optical to far infrared data, including deep medium-band infrared
photometry, hold great promise in selecting a more complete population
of $z\sim$ 3--4 galaxies \citep{spitler14}.  We have applied this
approach to select $z\sim$ 3--5 galaxies with expected escaping LyC
flux and report the results in a forthcoming companion paper.

As verified below, analyzing the colours and \lya\ EWs of $z\sim$ 3--4
galaxies with respect to the composite spectra expectations provides a
powerful tool to predict the fraction and wavelength extent of the
escaping ionizing photons.  In addition, we show that the analysis of
composite spectral samples on the $(U_n-G)$ vs.\ $(G-\cal{R})$ plane
has utility in estimating the overall escaping flux of $z\sim$ 3--4
photometric samples.

\subsection{Comparison to observed Lyman continuum
detections}\label{S01compare}

We compare the expectations of the composite spectra evolutionary
tracks with the results of previous spectroscopic LyC flux searches
using $U_nG\cal{R}$-selected LBGs.

\subsubsection{The Steidel et al. (2001) data set}\label{S01} 

\begin{figure}
\begin{center}
\scalebox{0.3}[0.3]{\rotatebox{90}{\includegraphics{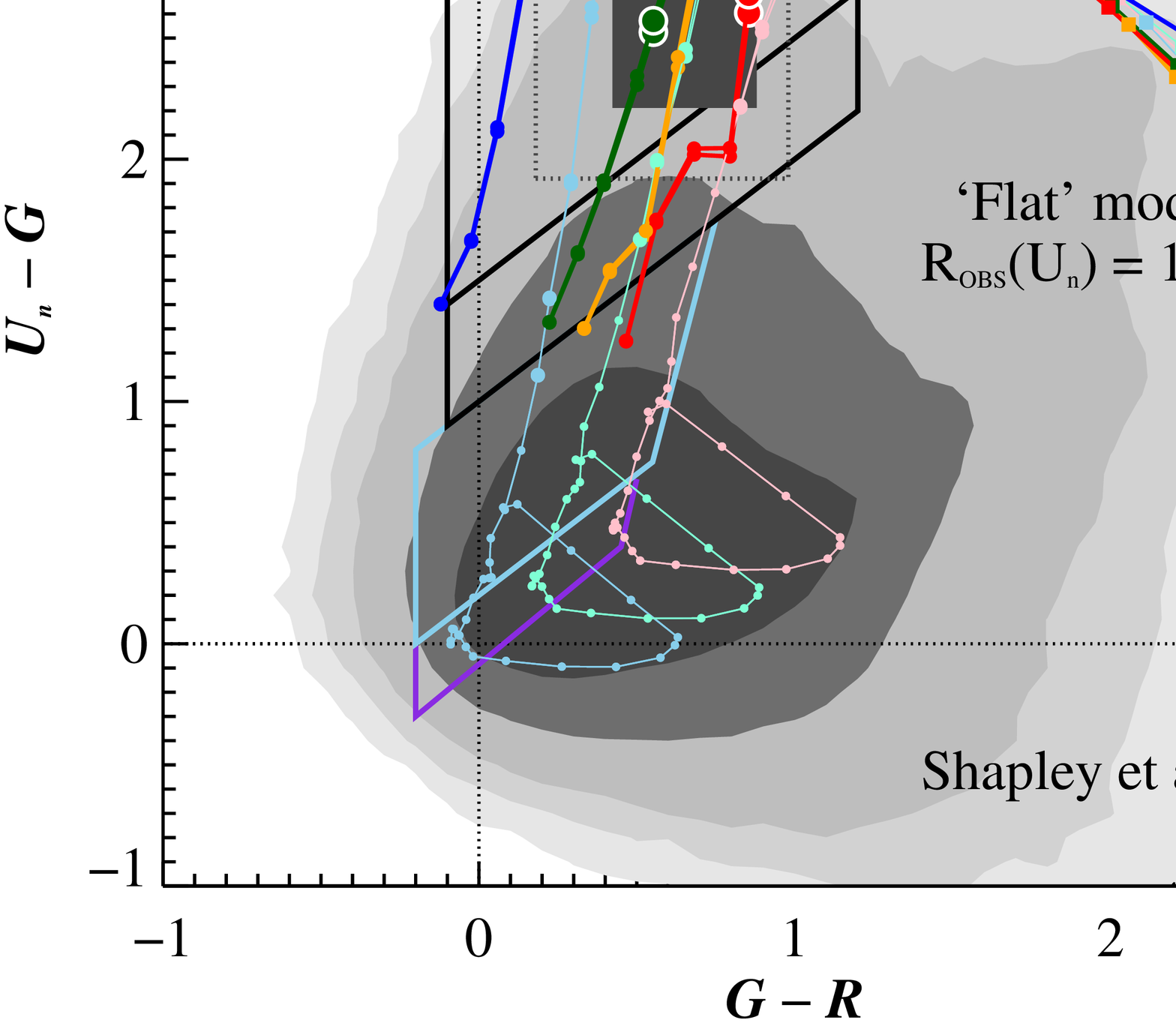}}}

\caption {\small Colour-colour plots similar to Figure~\ref{z3}.  {\it
Upper panel:} The 1$\sigma$ (black rectangle) and full range of
colours (dotted line window) for the 29 ${\langle
z\rangle=3.4\pm0.09}$ LBGs of \citet{steidel01} with
${\mbox{\robsl}=5.6}$ percent ${(\mbox{\fobs}=17.7)}$.  Overlaid are
the `step' and `flat' model evolutionary tracks for the four composite
spectra of \citet{shapley03}, quartiles q1--q4, shown as the red,
orange, dark-green, and dark-blue curves, respectively, and three
versions of a star-forming galaxy template with internal extinction of
${E(B-V)=0.0,0.15,}$ and 0.30 (light-blue, light-green, and pink
curves, respectively) applied using the \citet{calzetti00} law).  All
tracks have ${\mbox{\robsl}=5.6}$ percent.  Redshift steps of $\Delta
z=$ 0.1 are shown with coloured circles ${(z=2.7}$--3.5) and squares
${(z=3.6}$--4.5).  The large circles indicate ${z=3.4}$ for each
track.  The \citet{steidel01} LBGs have properties equivalent to the
quartile 4 LBGs (dark-blue tracks).  {\it Lower panel:} Similar to the
upper panel with the 1$\sigma$ and full range of colours for the 14
${\langle z\rangle=3.06\pm0.09}$ LBGs of \citet{shapley06} with an
average ${\mbox{\robsl}=1.7}$ percent ${(\mbox{\fobs}=58\pm25)}$ and
properties most similar to quartile 2 LBGs (orange tracks).  Here, the
1$\sigma$ range only includes objects with $U_n$ magnitudes brighter
than the limiting magnitude, thus, the true 1$\sigma$ range extends to
higher $(U_n-G)$.  The composite spectra and star-forming template
tracks have ${\mbox{\robsl}=1.7}$ percent, with large solid circles
marking ${z=3.06}$.  The large solid squares mark ${z=3.06}$ for
${\mbox{\robsl}=0.3}$ percent ${(\mbox{\fobs}=400}$, see text).  }

\label{S01S06}
\end{center}
\end{figure}

As discussed earlier, \citet{steidel01} measured the LyC flux of 29
$\langle z\rangle=3.40\pm0.09$ LBGs and reported an average
${\mbox{\fobs}=17.7\pm3.8}$ corresponding to
${\mbox{\robs(880\AA)}\sim 5.6}$ percent.  The sample has a full
redshift range of $z=$ 3.300--3.648 and $E(B-V)=$ -0.07--0.12.  The
upper panel in Figure~\ref{S01S06} plots the mean and 1$\sigma$
$(U_n-G)$ and $(G-\cal{R})$ values (dark rectangle) and full range of
values (dotted lines).  Overlaid in the upper panel are the composite
spectra evolutionary tracks for our two models using
${\mbox{\robsl}=5.6}$ percent.  In addition, we overlay a star-forming
galaxy template \citep{bruzual03} assigned ${\mbox{\robsl}=5.6}$
percent for the two models with three values of internal extinction
($E(B-V)$ = 0.0, 0.15, and 0.30) using the law of \citet{calzetti00}
and consistent with the presentation in S03.  Redshift intervals of
$\Delta z=$ 0.1 for the composite spectra and the templates are marked
(small circles for $z=$ 2.7--3.5 and small squares for $z=$ 3.6--4.5),
with large solid circles indicating $z=3.4$.

Composite spectra and the templates for $z=$ 3.4 objects with zero $<$
912\AA\ flux have infinite $(U_n-G)$ colours.  With this consideration
in mind, and the knowledge that the LBG sample was pulled from the
bluest quartile (i.e., their properties are most similar to quartile
4; the dark blue tracks), we find that the models predict the measured
LyC flux of the data very accurately.  In addition, the templates show
similar behaviour to the composite spectra which is a reassurance that
the movement in the $(U_n-G)$ vs.\ $(G-\cal{R})$ plane is not a
property specific to the composite spectra.  Over the full range of
colours for the data, the quartile 4 composite spectrum predicts a
redshift range of $z\sim$ 3.25--3.6, in close agreement with the data,
whereas we find $z\sim$ 3.0--3.6, $z\sim$ 3.0--3.5, and $z\sim$
3.05--3.3, for quartiles 3, 2, and 1, respectively.  The $E(B-V)$
values of the template are in agreement with the data (quartile 4 has
${E(B-V)=0.099\pm0.007}$) and show that galaxies with
${E(B-V)\gtrsim0.15}$ yield a poor fit.  Over the full data colour
range (dotted line window in Figure~\ref{S01S06}), we find that the
`step' model expectations for the quartile 4 composite spectrum ranges
from ${\mbox{\robs(800\AA)}\sim}$ 3--10 percent and \robsl\ $\sim$
2--7 percent for the `flat' model.

For $z\sim$ 3.4 galaxies, the $U_n$-band probes down to $\sim$740\AA\
and the LyC in the S03 spectra are measured from 880--910\AA.  The
direct agreement of the mean of the data with the `step' model shows
that the average escaping LyC flux is consistent with the integrated
results to $\sim$800\AA\ and the LyC flux can be, and likely is,
distributed over wavelengths shorter than 880\AA.  The `flat' model
shows that uniform LyC flux at the level found between 880--910\AA\
down to the $\sim$740\AA\ is within the range of the data but,
considering the `step' model results, suggests that the LyC flux does
not typically extend to wavelengths shortward of $\sim$800\AA\ for the
sample.

Finally, our analysis shows that the colours of the \citet{steidel01}
data set are unexpected for quartile 4 LBGs in the `standard' picture.
Although their $(G-\cal{R})$ colours are highly consistent with their
redshift and their spectral properties are in agreement, their
$(U_n-G)$ colours are much bluer than that predicted by composite
spectra (or star-forming templates) with little, or zero, LyC flux.
Instead, their colours are what is expected for the composite spectra
with an added LyC flux of $\sim$5\%.  Thus, application of our
technique on the \citet{steidel01} sample accurately predicts the
observed amount of LyC flux.

\subsubsection{Shapley et al. (2006) data set}\label{S06}

\citet{shapley06} obtain deep spectroscopy of 14 ${\langle
z\rangle=3.06\pm0.09}$ LBGs having colours and properties closer to
the mean of the full LBG population with overall properties closest to
quartile 2 LBGs.  The full redshift range of the sample is $z=$
2.76--3.30, with $E(B-V)=$ 0.03--0.19 (1$\sigma$ range), and
${\mbox{\fobs}=58\pm25}$, corresponding to \robsl\ $\sim$ 1.7 percent.
We note that \fobs\ $\sim$ 400, or \robsl\ $\sim$ 0.3 percent, when
excluding object C49 and D3 (see discussion in \S\ref{exp}).

The lower panel of Figure~\ref{S01S06} is plotted similarly to the
upper panel and compares the data to the composite spectra and
star-forming template expectations for \robsl\ = 1.7 percent.  The
large solid circles in the lower panel indicate $z=$ 3.06 for all
tracks.  The data average best matches the properties of quartile 2
LBGs and we find that the quartile 2 composite spectrum (orange
tracks) fits the best of the four quartiles.  We note that the
$(U_n-G)$ mean and 1$\sigma$ for the colour range plotted on
Figure~\ref{S01S06} is computed for the $U_n$-band detections and
lower limits (i.e., $U_n$ drop-outs), hence the true $(U_n-G)$ mean
and 1$\sigma$ is higher and broader than this range, however the
$(G-\cal{R})$ values remain the same.  Quartile 2 redshifts
distributed over the full range of the data colours are $z=$
2.95--3.3.  Quartile 1, 3, and 4 redshift ranges are $z=$ 2.85--3.15,
$z=$ 2.9--3.4, and $z=$ 3.0--3.35, respectively.  The range of
$E(B-V)$ values indicated by the template are in good agreement with
the data (quartile 2 has ${E(B-V)=0.136\pm0.006}$).  The `step' and
`flat' model for the quartile 2 composite spectrum produces \robsl\
$\sim$ 0--9 percent over the full data colour range.

We include the values for $z=$ 3.06 with \robsl\ = 0.3 percent, shown
using large squares (values for \robsl\ = 0.0 percent are nearly
identical) to assess the sample assuming negligible flux from objects
C49 and D3.  We find good agreement between the composite spectra
expectations and the data, especially when considering the true mean
and 1$\sigma$ colour dispersions when including $U_n$ drop-outs.  We
note that the composite spectra expectations are less sensitive to
differences in small amounts of observed LyC flux for the lower
redshift range but can confidently rule out moderate to high levels of
LyC flux.  Finally, we note that this sample was selected to be more
representative of the average properties of LBGs, including their
$(G-\cal{R})$ colours, but not necessarily their $(U_n-G)$ colours.
Whether or not the sample is representative of the average surviving
LyC flux of LBGs is unclear.  The colours of these galaxies are
expected for their redshifts and spectral type under the standard
approach and, as such, the sample is prone to small \robsl\ and is
expected to contain galaxies with little escaping LyC flux.

\begin{table*}
\begin{center}
\caption\normalsize{Lyman break galaxy observed and estimated
intrinsic Lyman continuum flux$^a$}\label{grid}
\begin{tabular}{lcccccccccccc}
\hline
Sample (model)$^b$ & \robs$(U_n)$ & \fescrel$^d$ & \fesc$^e$ & \robs$(U_n)$ & \fescrel$^d$ & \fesc$^e$ & \robs$(U_n)$ & \fescrel$^d$ & \fesc$^e$ \\
 & $lim^c$ & $lim^c$ & $lim^c$ & $lim+1.2^f$ & $lim+1.2^f$ & $lim+1.2^f$ & $99.0^g$ & $99.0^g$ & $99.0^g$ \\
\hline
Quartile 1 (Flat) & 3.0$^{+0.8}_{-0.3}$ & 21.9$^{+5.6}_{-2.3}$   &  4.4$^{+1.1}_{-0.5}$ & 3.0$^{+0.8}_{-0.3}$ & 21.3$^{+5.7}_{-2.1}$   &  4.3$^{+1.2}_{-0.4}$ & 2.9$^{+0.7}_{-0.3}$ & 20.8$^{+5.4}_{-1.9}$ &  4.2$^{+1.1}_{-0.4}$\\
Quartile 2 (Flat) & 3.4$^{+0.9}_{-0.6}$ & 24.1$^{+6.8}_{-4.4}$   &  6.7$^{+1.9}_{-1.2}$ & 3.3$^{+0.9}_{-0.6}$ & 23.4$^{+6.3}_{-4.1}$   &  6.5$^{+1.9}_{-1.7}$ & 3.2$^{+1.1}_{-0.6}$ & 22.8$^{+6.2}_{-4.1}$ &  6.3$^{+1.7}_{-1.1}$\\
Quartile 3 (Flat) & 4.3$^{+1.0}_{-0.9}$ & 30.9$^{+7.2}_{-6.5}$   & 10.0$^{+2.3}_{-2.1}$ & 4.1$^{+0.9}_{-0.9}$ & 29.9$^{+6.8}_{-6.3}$   &  9.6$^{+2.2}_{-2.0}$ & 4.0$^{+0.9}_{-0.8}$ & 28.6$^{+6.6}_{-5.9}$ &  9.2$^{+2.1}_{-1.9}$\\
Quartile 4 (Flat) & 7.1$^{+2.4}_{-2.4}$ & 51.1$^{+17.6}_{-17.5}$ & 20.1$^{+6.9}_{-6.9}$ & 6.8$^{+2.2}_{-2.3}$ & 48.8$^{+16.4}_{-16.7}$ & 19.2$^{+6.4}_{-6.6}$ & 6.5$^{+2.2}_{-2.2}$ & 46.8$^{+16.0}_{-16.0}$&18.4$^{+6.3}_{-6.3}$\\
Average (Flat)    & 4.3$^{+1.3}_{-1.0}$ & 31.3$^{+9.0}_{-7.3}$   &  9.1$^{+2.6}_{-2.1}$ & 4.2$^{+1.2}_{-1.0}$ & 30.2$^{+8.6}_{-6.9}$   &  8.8$^{+2.5}_{-2.0}$ & 4.0$^{+1.2}_{-0.9}$ & 29.1$^{+8.3}_{-6.6}$ &  8.4$^{+2.4}_{-1.9}$\\
Average (Flat) corr.$^h$& 3.2$^{+0.6}_{-0.6}$ & 23.1$^{+4.3}_{-4.4}$ &  6.7$^{+1.2}_{-1.3}$ & 3.4$^{+0.7}_{-0.7}$ & 24.4$^{+4.8}_{-4.8}$   &  7.1$^{+1.4}_{-1.4}$ & 3.0$^{+0.6}_{-0.6}$ & 21.9$^{+4.0}_{-4.0}$ &  6.4$^{+1.2}_{-1.2}$\\
\hline
Quartile 1 (Step) & 3.3$^{+0.7}_{-0.0}$ & 23.8$^{+4.8}_{-1.9}$   &  4.8$^{+1.0}_{-0.4}$ & 3.2$^{+0.7}_{-0.0}$ & 23.2$^{+5.0}_{-0.0}$   &  4.7$^{+1.0}_{-0.0}$ & 3.1$^{+0.6}_{-0.0}$ & 22.6$^{+4.7}_{-0.1}$ &  4.6$^{+1.0}_{-0.0}$\\
Quartile 2 (Step) & 4.0$^{+0.6}_{-0.1}$ & 28.9$^{+4.3}_{-0.4}$   &  8.0$^{+1.2}_{-0.1}$ & 3.9$^{+0.5}_{-0.0}$ & 28.0$^{+3.8}_{-0.3}$   &  7.8$^{+1.1}_{-0.1}$ & 3.8$^{+0.5}_{-0.0}$ & 27.2$^{+3.8}_{-0.3}$ &  7.5$^{+1.1}_{-0.1}$\\
Quartile 3 (Step) & 5.4$^{+0.5}_{-0.2}$ & 38.7$^{+3.7}_{-1.3}$   & 12.5$^{+1.2}_{-0.4}$ & 5.2$^{+0.4}_{-0.2}$ & 37.3$^{+3.4}_{-1.4}$   & 12.0$^{+1.1}_{-0.5}$ & 4.9$^{+0.5}_{-0.2}$ & 35.5$^{+3.6}_{-1.6}$ & 11.4$^{+1.2}_{-0.5}$\\
Quartile 4 (Step) & 8.6$^{+2.3}_{-1.8}$ & 62.6$^{+16.8}_{-13.2}$ & 24.6$^{+6.6}_{-5.2}$ & 8.2$^{+2.3}_{-1.8}$ & 59.4$^{+16.3}_{-13.1}$ & 23.3$^{+6.4}_{-5.1}$ & 7.8$^{+2.2}_{-1.8}$ & 56.2$^{+16.0}_{-12.6}$&22.1$^{+6.3}_{-4.9}$\\
Average (Step)    & 5.2$^{+1.0}_{-0.4}$ & 37.5$^{+7.2}_{-2.9}$   & 10.9$^{+2.1}_{-0.8}$ & 5.0$^{+1.0}_{-0.4}$ & 36.0$^{+6.9}_{-2.8}$   & 10.5$^{+2.0}_{-0.8}$ & 4.8$^{+0.9}_{-0.4}$ & 34.5$^{+6.8}_{-2.6}$ & 10.0$^{+2.0}_{-0.8}$\\
Average (Step) corr.$^h$& 3.9$^{+0.4}_{-0.3}$ & 28.1$^{+2.9}_{-1.8}$ &  8.2$^{+0.8}_{-0.5}$ & 4.1$^{+0.5}_{-0.3}$ & 29.7$^{+3.3}_{-2.2}$   &  8.6$^{+1.0}_{-0.6}$ & 3.7$^{+0.4}_{-0.2}$ & 26.3$^{+2.7}_{-1.6}$ &  7.6$^{+0.8}_{-0.5}$\\
\hline
\end{tabular}
\end{center}
$^a${All values are in percentages}\\
$^b${The quartiles of LBGs and full LBG sample are taken from
\citet{shapley03}.}\\
$^c${$U_n$-band non-detections set to the 1$\sigma$ limiting magnitude
of the field}\\
$^d${\fescrel\ is determined using an estimated IGM attenuation
factor, exp $(\tau^{\rm IGM}_{\rm LyC})$ = 2.4 for $z=$ 3.0 and
$(L_{1500}/L_{900})^{\rm int}=$ 3.0}\\
$^e${\fesc\ = \fescrel\ $\cdot$ \fescfif, where \fescfif\ is
computed using the $E(B-V)$ values for each respective quartile.}\\
$^f${Intermediate value between $U_n(lim)$ and $U_n=99.0$ and
equivalent to $U_n$-band non-detections set to the 1$\sigma$ limiting
magnitude + 1.2}\\
$^g${$U_n$-band non-detections set to $U_n=99.0$}\\
$^h${Values corrected for $z<$ 2 objects, AGN, and line-of-sight
contamination {\it (see text)}}
\end{table*}

\subsection{The escape fraction of Lyman break galaxies}\label{lcgs}

We have demonstrated that the redshift, colours, and observed LyC flux
of small samples of LBG spectra are predictable using the modified
composite spectra using {\it only} photometric and \lya\ EW
information.  The evolutionary tracks of the modified composite
spectra have further utility in measuring the observed average LyC
flux of the full LBG population when applied to statistically
significant spectroscopic samples.  In addition, the colour
distribution of composite spectral samples with accurate photometric
uncertainties can be used to estimate the average observed LyC flux
for LBG photometric samples.  However, it must be kept in mind that
our approach estimates the average LyC flux acquired via spectroscopy
or within photometric apertures and will not include LyC flux that may
exist offset from the UV continuum \citep[e.g.,][]{iwata09,nestor13}
in which spectroscopic slits are typically aligned and/or LyC flux
outside the area of the slit.  Below, we estimate the average observed
LyC flux of the $z\sim$ 3 LBG population by matching the colours and
\lya\ EWs of individual S03 spectra with the composite spectra
predictions.

\subsubsection{The average Lyman continuum flux from \zzz\ Lyman break
galaxies}

We compare the net \lya\ EW and colours of the 794 LBG spectra of
\citet{shapley03} to a grid of composite spectra evolutionary tracks
to measure the LyC flux on an individual galaxy basis.  The grid
includes the tracks shown in Figure~\ref{z3}, but contains many more
tracks sampling $>10\times$ higher resolution of \robsl\ values.  The
individual spectra are matched to their respective composite spectrum
track based on their \lya\ EW and assigned the \robsu\ of the closest
track value based on their colours.  We use the results from this
statistical sample to estimate the average LyC flux of \zzz\ LBGs as
determined by their photometry.

The $(G-\cal{R})$ uncertainties are $\lesssim$0.15 magnitudes, thus
there is a small dispersion of the individual galaxies about the
redshift expectations of their respective composite spectrum
evolutionary track.  In addition, there is a formal $(U_n-G)$ $\sim$
0.25 magnitude uncertainty that increases to $(U_n-G)$ $\sim$ 0.6--0.7
magnitudes for objects near the magnitude limit (we consider $U_n$
drop-outs below).  Finally, there can be an uncertainty in individual
\lya\ EW measurements of $\sim$30--50 percent \citep{shapley03} which
would affect which composite spectral track the LBGs are assigned, in
particular for border-line cases.  As a result, the values for
individual cases are rather coarsely estimated, however, the $\sim$200
galaxies in each quartile and the combined values of the 794 LBGs in
the four quartiles provide us with an accurate average measure of the
LyC flux of \zzz\ LBGs meeting the standard colour-selection criteria.

When measuring the individual spectra, we assign $U_n$ drop-outs
either the limiting magnitude of the field to provide an upper limit
to the true level of observed LyC flux or $U_n=$ 99.0 to provide a
lower limit.  The results from matching the data to the `flat' and
`step' model grids are listed in Table~\ref{grid}.

Comparing the data to the `flat' model yields an average \robsu\ =
4.3$^{+1.3}_{-1.0}$ and 4.0$^{+1.2}_{-0.9}$ percent for $U_n$
magnitude limit and $U_n=$ 99.0 drop-out assignments, respectively.
To determine how these values relate to the intrinsic escape fraction
of LyC photons from \zzz\ LBGs, we refer to equation~\ref{fesc-def}
which shows that \fesc\ = \fescrel\ $\cdot$ \fescfif.  We obtain
\fescrel\ by correcting \robsu\ by an estimated $z=$ 3.0 average IGM
attenuation factor, exp $(\tau^{\rm IGM}_{\rm LyC})$ = 2.4
\citep[e.g.,][]{shapley06}, and a standard assumption of
$(L_{1500}/L_{900})^{\rm int}=$ 3.0.  We use the extinction law of
\citet{calzetti00} and mean $E(B-V)$ values for each quartile to
determine \fescfif.  These assumptions produce \fescrel\ =
31.3$^{+9.0}_{-7.3}$ and 29.1$^{+8.3}_{-6.6}$ percent and \fesc\ =
9.1$^{+2.6}_{-2.1}$ and 8.4$^{+2.4}_{-1.9}$ percent for the two $U_n$
drop-out assignments, respectively.

The `step' model produces an average \robsu\ = 5.2$^{+1.0}_{-0.4}$ and
4.8$^{+0.9}_{-0.4}$ percent; \fescrel\ = 37.5$^{+7.2}_{-2.9}$ and
34.5$^{+6.8}_{-2.6}$ percent; and \fesc\ = 10.9$^{+2.1}_{-0.8}$ and
10.0$^{+2.0}_{-0.8}$ percent for the $U_n$ magnitude limit and $U_n=$
99.0 drop-out assignments, respectively.  There are small, or no,
differences between the `flat' and `step' models on the $(U_n-G)$
vs. $(G-\cal{R})$ plane for $z\lesssim$ 3.2 LBGs, but the differences
become significant at higher redshifts.  Given that the `step' model
provides a potentially better fit to the reported LBG \fescrel\
measurements in the literature, especially for the \citet{steidel01}
data at $z=$ 3.4, we report the `step' model values.  Moreover, we
adopt \robs$(U_n)=$ 5.0$^{+1.0}_{-0.4}$ and \fesc\ =
10.5$^{+2.0}_{-0.8}$ percent for \zzz\ LBGs, which are intermediate to
the $U_n$ magnitude limit and $U_n=$ 99.0 assignment values.  We find
that these values also equate to \robs$(U_n)$ when assigning $U_n$
drop-outs an average value 1.2 magnitudes fainter than their field
1$\sigma$ limiting magnitudes.  The results for the intermediate case
for both models are listed in Table~\ref{grid} for comparison.

\subsubsection{Lyman break galaxy \robsu\ distributions}

\begin{figure}
\begin{center}
\scalebox{0.19}[0.19]{\rotatebox{90}{\includegraphics{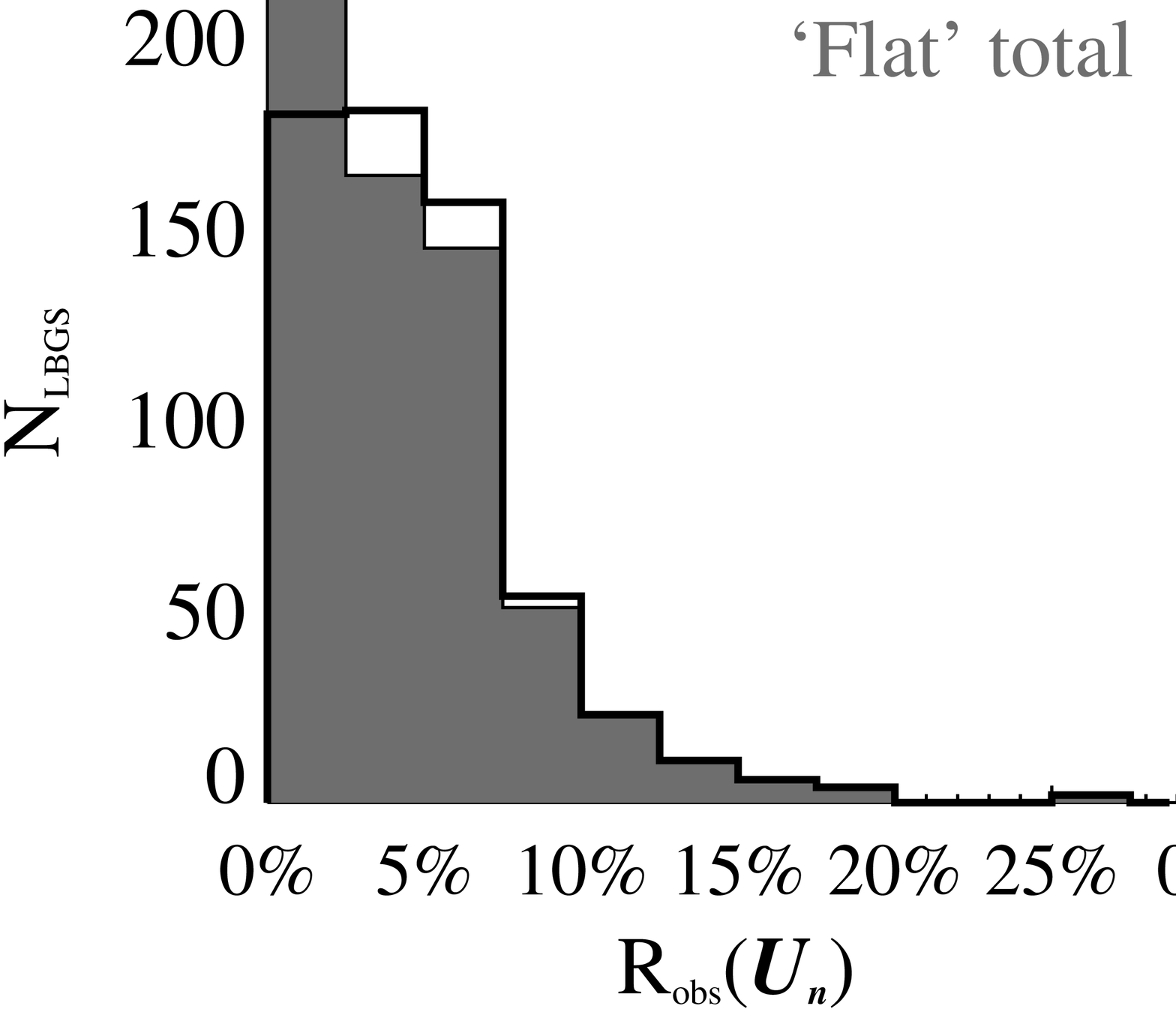}}}

\caption {\small Histograms of the ratio of observed LyC in the
$U_n$-band to UV continuum flux at 1500\AA, \robsu, for \zzz\ Lyman
break galaxies (LBGs).  Plotted from top to bottom are the \robsu\
distributions for quartiles 1--4 and the total LBGs of
\citet{shapley03} with the `flat' and `step' models labelled.  In each
panel, the black and coloured histograms denote the values when
assigning $U_n$-band non-detections (drop-outs) the $U_n$ limiting
magnitude of the field and $U_n=$ 99.0, respectively.  }

\label{robs}
\end{center}
\end{figure}

Histograms of the \robsu\ distributions for the two models and the
separate quartiles are presented in Figure~\ref{robs}.  The
distributions differ and the data imply that LBGs with stronger net
\lya\ in emission have higher \robsu.  One explanation for this
relationship is the restriction of galaxies to the standard
colour-selection region.  The bluer $(G-\cal{R})$ colours of LBGs with
properties similar to quartile 3 and 4 are able to have higher \robsl\
and still meet the standard criteria, whereas LBGs with properties
similar to quartile 1 and 2 with moderate to high \robsu\ will fall
out and their distribution will be truncated.  An important result of
the \robsu\ distributions is that the truncations demonstrate that
there exists a significant fraction of galaxies outside the standard
criteria and that those galaxies have higher \robsl\ and \fesc.

We perform two-sided Kolmogorov-Smirnov tests to quantify the spectral
type \robsu\ distributions and report the results in Table~\ref{KS}.
Neighbouring spectral types show a small to good probability of being
pulled from the same parent population ($p\sim$ 0.02 --0.78) except
for quartile 3 and quartile 4 LBGs.  These latter two spectral samples
have K--S probabilities of 10$^{-5}$ to 10$^{-6}$ indicating a more
dramatic change in behaviour of the quartile 4 LBGs with the others.
Further solidifying this result are the K--S test probabilities
ranging from $\sim$10$^{-11}$ to 10$^{-23}$ for quartile 4 LBGs when
tested with quartile 1 and quartile 2.  Nearly all quartile 4 LBGs
would be detected as LAEs in narrowband surveys. Because LAEs are a
natural extension to the quartile 4 LBG population
\citep[e.g.,][]{cooke09,cooke13}, the indication of a higher LyC
escape fraction from quartile 4 LBGs/LAEs compared to other LBG
quartiles is consistent with LyC measurements of \zzz\ LAEs compared
to LBGs \citep{nestor13}.

The \robsu\ results were assigned to the composite spectral samples
with the photometric corrections discussed in \S\ref{conv_exp} (i.e.,
a conservative addition of a $U_n$ redleak, spectroscopic slit flux
loss, and $U_n$ drop-outs set to the $U_n$ limiting magnitude) and
compared to the photometric data.  To estimate the range of observed
or predicted \robsu, each sample was taken as a whole or divided into
multiple parts.  Overall, any value of \robsu\ from zero to $\sim$5
percent produces a distribution that has the general form of the S03
data and consists of a significant fraction of galaxies that extends
outside the standard region.  As the value of \robsu\ is increased,
the $(U_n-G)$ mean of the distribution ranges from approximately that
shown in the upper panel of Figure~\ref{conventional} to approximately
that of the data (Figure~\ref{conventional}, bottom panel).

To test whether the truncations seen in Figure~\ref{robs} are due
entirely to the standard colour-selection criteria or whether some
fraction of the relationship is intrinsic, the subsets of the
composite spectral samples were given equivalent \robsu\ values and
tested at successively higher values.  The resulting distributions do
not reflect the form or mean of the data in the sense that, as the
value of \robsu\ is increased, the distributions shift to smaller
$(G-\cal{R})$ values as they moves to smaller $(U_n-G)$ values.
Constant values of \robsu\ $\gtrsim$10 percent evenly assigned to each
quartile are ruled out.  However, good matches to the data
distribution only persist when we assign a range of \robsu\ values
that are dependent on quartile consistent with that found for the
data, such that quartiles with stronger \lya\ emission have higher
values of \robsu.  The results suggest that part of the relationship
between the level of observed LyC flux and the observed strength of
\lya\ emission may be intrinsic.

\begin{table}
\begin{center}
\caption\normalsize{LBG spectral type \robsu\ K-S test results$^a$}\label{KS}
\begin{tabular}{lcccc}
\hline
Samples$^b$ & Flat (lim)$^c$ & Flat (99)$^d$ & Step (lim)$^c$ & Step (99)$^d$\\
\hline
Q1, Q2 & 6.4$\times$10$^{-1}$ & 7.8$\times$10$^{-1}$ & 1.6$\times$10$^{-1}$ & 2.1$\times$10$^{-1}$\\
Q1, Q3 & 4.6$\times$10$^{-3}$ & 1.1$\times$10$^{-2}$ & 2.9$\times$10$^{-5}$ & 1.6$\times$10$^{-4}$\\
Q1, Q4 & 3.9$\times$10$^{-14}$ & 7.7$\times$10$^{-12}$ & 3.7$\times$10$^{-23}$ & 7.1$\times$10$^{-19}$\\
Q2, Q3 & 1.0$\times$10$^{-1}$ & 1.4$\times$10$^{-1}$ & 2.2$\times$10$^{-2}$ & 3.3$\times$10$^{-2}$\\
Q2, Q4 & 4.8$\times$10$^{-13}$ & 1.8$\times$10$^{-11}$ & 2.1$\times$10$^{-16}$ & 6.0$\times$10$^{-13}$\\
Q3, Q4 & 1.5$\times$10$^{-6}$ & 9.8$\times$10$^{-6}$ & 2.0$\times$10$^{-7}$ & 9.7$\times$10$^{-6}$\\
\hline
\end{tabular}
\end{center}
$^a${Two-sided Kolmogorov-Smirnov probability that the two samples are
pulled from the same parent population.}\\
$^b${Samples Q1--Q4 refer to quartile 1--4}\\
$^c${$U_n$-band non-detections set to the 1$\sigma$ limiting magnitude
of the field}\\
$^d${$U_n$-band non-detections set to $U_n=99.0$}
\end{table}

\subsubsection{Contamination to \robsu}\label{contam}

Foreground (lower redshift) objects in the line of sight to ${z\sim3}$
LBGs can introduce restframe non-ionising flux that could contaminate
LyC flux measurements.  In fact, the significant LyC flux observed for
one of the LBGs in the sample of \citet{shapley06} is likely longer
wavelength flux from a lower redshift galaxy \citep[cf.,][]{nestor13}
resulting in $\sim$7\% contamination.  In addition, \citet{nestor13}
find line-of-sight contamination for the LBGs in their narrowband
survey with LyC detections may be as high as 62\%, however, 38 LBGs in
their sample of 41 have no detectable flux, thus the overall LBG
contamination from line-of-sight sources that are capable of
contributing measurable flux (to m $\sim28$) to the $U_n$-band is
likely $\sim$7 percent.

We remark that our composite spectra are constructed from the LBG
spectra of S03.  As such, they accurately represent the colours of
\zzz\ LBGs selected using standard colour selection criteria longward
of the Lyman limit.  Because we replace all flux shortward of the
Lyman limit with artificial flux values, the composite spectra and the
composite spectral sample distributions are not subject to
line-of-sight contamination.  Only the actual data of S03 on an
individual object basis are subject to contamination, and thus our
\zzz\ LBG \robsu\ and \fesc\ estimates.  Below, we estimate the
fraction of \zzz\ LBGs in the S03 sample that have contamination to
their $U_n$-band photometry by foreground line-of-sight sources down
to the faintest level which would affect our LyC flux measurements,
$U_n\sim28$.  However, first we review a few relevant photometric
details.

In the S03 survey, the $U_nG\cal{R}$ images were aligned/registered
and then smoothed with a Gaussian kernel to the same seeing FWHM.
Then LBGs are identified in the $\cal{R}$-band image by their
$\sim$1500\AA\ flux in isophotal apertures with $\sim$2 arcsec
diameters, which were then grown to $\sim$3 arcsec ($\sim$0.1-0.15
magnitude difference between the two apertures).  The pixels defining
the extent of the $\cal{R}$-band detections for each LBG were then
used to determine the flux in the other filters.  This was done
because many of the objects are not detected in the $U_n$-band.  The
half-light radii of \zzz\ LBGs range from $\sim$1--5 kpc, or
$\sim$0.1--0.6 arcsec \citep{ferguson04}.  Because the seeing FWHM for
the observations ranged from 0.7--1.2 arcsec, \zzz\ LBGs appear as
point sources or near point sources.  As as result, lower redshift
objects must overlap the point source-like area of the LBGs in the
$U_n$ filter if they are to contribute to the LyC flux.

This constraint is one important difference between our measurement
and narrowband LyC imaging searches.  We are not measuring the
extended and/or offset LyC flux that is commonly observed in those
surveys.  Extended and/or offset flux increases the probability of
lower redshift source contamination and source confusion by probing an
additional area with respect to the stellar component of the LBG.
However, because we do not include extended or offset regions, we
measure a lower limit to the true amount of escaping LyC flux but
potentially a higher value than that measured through a spectroscopic
slit.

Low redshift sources have a large range in magnitude and sizes/shapes
and a high density in the deep images, whereas S03 LBGs are small
sources with $\cal{R}$ = 19.0--25.5 and a relatively low surface
density of $\sim$1.8 LBGs arcmin$^{-1}$.  Thus, it is much more
accurate to simulate the overlap of field sources on the areas of LBGs
than vice-versa.  Lower redshift objects in the line of sight that are
$\sim$1--4 magnitudes fainter in the $U_n$-band than the $G$-band for
a given LBG could measurably contribute to the LyC flux.  Finally,
brighter objects that are not completely extracted or that pose
problematic blending with the LBGs can affect the $U_nG\cal{R}$
photometry in a way to push the colours outside the standard selection
region.

We use the 5-year stacked, wide-field $u^*g'r'i$ images of the
Canada-France-Hawaii Telescope Legacy Survey (CFHTLS) Deep
Fields\footnote{General information for the CFHTLS Deep fields and
data products can be found at:
www.cfht.hawaii.edu/Science/CFHLS/cfhtlsdeepwidefields.html and the
associated links} to estimate the fraction of LBGs with line-of-sight
contamination.  The $u^*$ filter is similar to, but has a redder
bandpass than, the $U_n$ filter, the $g'$ and $G$ bands are similar,
and the $\cal{R}$ filter falls in-between the $r'$ and $i'$ filters.
The CFHTLS has a limiting magnitude of $u^*$ = 27.4, signal-to-noise
ratio (S/N) = 5.  If we include all objects with ${\mbox S/N}\ge3$, we
probe depths to roughly $u^*\sim$ 28.0, similar to the 1$\sigma$
limits of the S03 survey.  The CFHTLS images provide a deep test of
the contamination and the results are conservative in the sense that
most field objects are brighter in $u^*$ as compared to the bluer
$U_n$ bandpass.

We generate simulated LBGs using the composite spectra technique which
uses magnitudes from the $\cal{R}$-band magnitude distribution and
computes their $U_n$ and $G$ magnitudes assuming zero LyC flux.  We
then compute the $u^*g'r'i'$ magnitudes for the LBGs and place them in
the CFHTLS square degree images with the surface density and area
equivalent to S03 survey fields.  As a control, we lay the same
simulated LBGs in a ``blank'' image with simulated sky noise at the
level measured in the CFHTLS images.

The LBGs are then extracted using $SExtractor$ \citep{ba96} in a
consistent manner to S03.  We then compare object magnitudes from the
``blank'' and real images to the faint limit where we find equal
numbers of positive and negative flux enhancements as a result of sky
fluctuations.  Objects that had their magnitudes enhanced to the point
where they were pushed out of the standard selection region are not
included.  We measure a 5.3\% net positive enhancement to the
$u^*$-band flux of objects in the standard LBG selection region.  As
an additional test, we multiply the $SExtractor$ segmentation map for
the LBGs in the control field with the $u^*$-band segmentation map
which uses $i'$-band defined apertures (to mimic S03 $\cal{R}$-band
apertures).  We find detection of $u^*$-band flux in 5.2\% of the
apertures.  As a result, we adopt a 5.3\% contamination fraction to
the S03 sample from lower redshift line-of-sight sources.

Finally, for a total contamination to the LyC flux from all sources,
we include an estimate of $z<$ 2 sources and AGN contained in the S03
data.  From their 940 spectra, 40 Galactic stars, five $z<$ 2
galaxies, and 28 AGN were identified.  Combining this 7.8\%
contamination with the line-of-sight contamination, we estimate a
total 13.1\% contamination to the photometric sample.  Assuming that
the contaminants make up the 13.1\% of the galaxies that have the
highest \robsu, we remove these objects from the data and produce the
corrected average `flat' and `step' model values listed in
Table~\ref{grid}.

We conclude that foreground and AGN contamination are not the main
contributors to the $U_n$-band flux excess observed in the S03 data
distribution (Figure~\ref{conventional}, lower panel) and that genuine
LyC flux remains the best candidate to explain the bulk of the excess.


\section{Lyman continuum galaxies}

The modified composite spectra evolutionary tracks and \robsu\
distribution truncations, the colour-colour distribution of the data,
and the spectroscopic confirmations of the VVDS all point to the
existence of a population of galaxies outside the standard criteria
with bluer $(U_n-G)$ colours.  Because the evolutionary tracks of the
composite spectra indicate that these galaxies have moderate to
significant amount of LyC flux, and that essentially none should have
zero LyC flux, we refer to galaxies in this region as Lyman continuum
galaxies (LCGs).

\begin{figure}
\begin{center}
\scalebox{0.29}[0.29]{\rotatebox{90}{\includegraphics{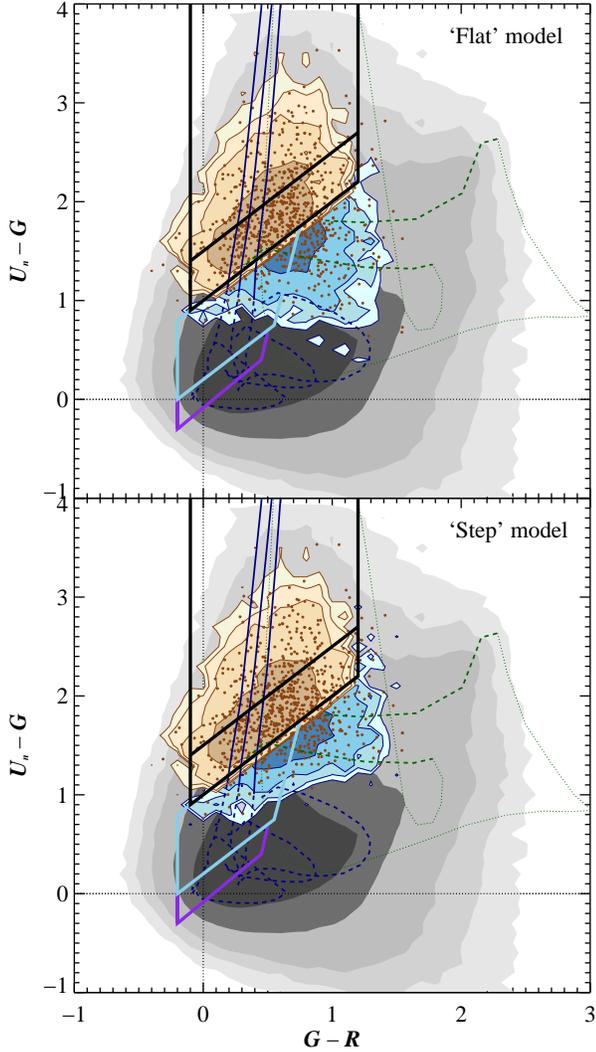}}}

\caption {\small Colour-colour plot similar to
Figure~\ref{conventional} illustrating a fit of the composite spectral
samples to the distribution to the photometric data of
\citet{steidel03}. The grey contours are the typical colours of field
objects.  The combined coloured contours overlaid reflect the
continuous distribution of four composite spectra samples for the
`flat' model (upper panel) and `step' model (lower panel) with LyC
escape values discussed in the text.  Galaxies meeting the standard
LBG colour selection criteria are shown with tan/brown contours and
those outside the standard criteria (Lyman continuum galaxies; LCGs)
are shown with blue contours.  The brown filled circles are randomly
selected composite spectra values for the full sample.  This match to
the data results in LCGs comprising $\sim$30--33 percent of \zzz\
galaxies.  The range of colours probed by LCGs is in very good
agreement with the colour distribution of \zzz\ galaxies
spectroscopically confirmed in the literature.  }

\label{fesc}
\end{center}
\end{figure}

The full behaviour of $z=$ 2.7--3.5 galaxies can be visually
reproduced by extending the maximum range of the quartile \robsu\
values and keeping the quartile-to-quartile ratio (i.e., countering
the effect of truncation).  Figure~\ref{fesc} presents one such
realisation when extending the measured quartile \robsu\ values in
Table~\ref{grid} by a factor of 1.5.  All values of \robsu\ in
Table~\ref{grid} produce similar distributions, along with
multiplicative factors ranging from 1--2.5.  The match in
Figure~\ref{fesc} shows that LCGs make up $\sim$32 percent of the full
$z\sim$ 3 galaxy population, or $\sim$50\% the number of LBGs.
Interestingly, the fraction of LCGs is only a few percent lower
without {\it any} extension of the \robsu\ values.  If these
realisations are representative of the true galaxy distribution, the
current census of UV-luminous \zzz\ star-forming galaxies needs to be
increased by a factor of $\sim$1.5.  Moreover, LCGs are expected to
have high levels of observed LyC flux and may contribute as much total
ionising flux as the LBG population or more.

Statistical fits do not yield unique solutions to the distribution
fits because of the unknown LCG distribution.  As a result, our simple
\robsu\ distribution prescription in Figure~\ref{fesc} is not a formal
fit, but is meant to help illustrate the $(U_n-G)$ vs. $(G-\cal{R})$
location of \zzz\ LCGs and estimate their number density.
Nevertheless, our attempts to mimic the photometric distribution of
the data using a wide range of \robsu\ values, and distributions of
the values, have uncovered some interesting results and unavoidable
constraints for each quartile, and for the full population, that are
reinforced by the distribution and LyC measurements of the 794
spectra.

Firstly, no combination of LyC flux assignments to the composite
spectral samples, including zero LyC flux, can achieve a similar
distribution as the data, and/or a peak density near the $(U_n-G)$ =
$(G-\cal{R})$ + 1 boundary, without incurring a significant fraction
of galaxies outside the standard LBG selection region.  As discussed
in \S\ref{conv_exp}, this fact is also revealed by the data
distribution.  The location on the colour-colour plane of the galaxies
predicted outside the standard selection region is highly consistent
with the colours of the spectroscopically confirmed VVDS $z\sim$ 3
galaxies found in that region \citep{lefevre05b,paltani07}.

Secondly, the galaxies predicted outside the standard colour-selection
region are expected to have moderate to high levels of observed LyC
flux (1) from our composite spectral analysis and (2) because they
have $(G-\cal{R})$ colours that are typical for their redshifts but
bluer than typical $(U_n-G)$ colours, denoting an excess of $U_n$-band
flux.  The $U_n$-band probes the restframe LyC at $z\gtrsim$ 3 and any
excess $U_n$-band flux for a $z\gtrsim$ 3 galaxy equates to excess LyC
flux.

Thirdly, to match the shape and location of the data distribution,
\robsl\ must be assigned as a function of \lya\ EW (quartile) such
that LBGs with strong \lya\ in emission (quartile 4) have higher
observed LyC flux than those with successively weaker \lya\ emission.
This relationship is seen for the individual spectra by quartile in
Table~\ref{grid} for both models and all $U_n$ magnitude assignments.
From our \robsu\ assignment tests, the relationship appears to exist
beyond the effect caused by the constraint of galaxies to the standard
colour-selection region.  A positive correlation between \lya\ and
escaping LyC flux is expected to first order (after the net loss in
\lya\ to LyC photons) given that the \lya\ and LyC radiation transfer
are both sensitive to the presence of H\textsc{i} in the ISM.  Thus it
may be that galaxies with \lya\ in emission and their corresponding
properties, such as bluer continua, weaker ISM metal-line features,
compact morphology, field-like environments, and potentially higher
merger rates, or their orientation may cause higher intrinsic LyC
escape fractions.

Finally, the composite spectra evolutionary tracks indicate that a
fraction of $z\lesssim$ 3.0 LCGs, comprised largely of quartile 4
galaxies, fall in the standard `BX' selection region, whereas
$z\gtrsim$ 3.0 LCGs (from all quartiles) reside outside any standard
selection region.  Essentially all quartile 1 and 2 LCGs with \robsl\
$\gtrsim$ 10 percent are missed using standard selection criteria and
all $z>3.5$ LCGs from all quartiles with \robsl\ $\gtrsim$ 1 percent
are missed using standard \zzzz\ criteria.

\subsection{Estimating the average Lyman continuum flux from Lyman
continuum galaxies}

A rough estimate of the mean observed LyC flux of LCGs can be gleaned
from our match of the composite spectral samples to the data
distribution.  A series of input \robsu\ values for both models that
match the data distribution produce similar results and return an
\robsl\ = 15.8 $\pm3.3$ percent.  Using the same assumptions above for
LBGs, this value corresponds to \fesc\ = 33.0 $\pm6.9$ percent.  Both
values are theoretical and reflect the observed LyC flux after
contamination correction.  Using this approach on the combined set of
$z\sim$ 3 galaxies, we find \robsl\ = 7.6 $\pm1.7$ percent and \fesc\
= 15.9 $\pm3.5$ percent.  These values are model dependent and in
\S\ref{exp} we discuss that the intrinsic value of $L_{1500}/L_{900}$
may vary between 1--5.5.  In addition, it may be the case that LCGs
are found along lines of sight with the fewer than average intervening
absorption systems and thus the attenuation by the IGM may be lower.


\section{Summary}\label{conc}

The $z\sim$ 3--4 Lyman break technique preferentially selects galaxies
with little or no LyC flux.  A re-examination of the distribution of
\zzz\ LBGs on the $(U_n-G)$ vs. $(G-\cal{R})$ plane indicates that a
significant fraction of \zzz\ galaxies reside outside the standard LBG
colour selection region.  To assess the completeness of $z\sim$ 3--4
LBG colour selection and to better understand LBG LyC flux measurement
results in the literature, we perform spectrophotometry on composite
spectra formed from 794 $z\sim$ 3 LBGs and apply various levels of
escaping ionising ($<$ 912\AA) flux.  We find the following results
from the behaviour of the evolutionary tracks of the modified
composite spectra and the colour distributions of the composite
spectral samples that include photometric uncertainties.

\begin{enumerate}

\item Spectrophotometry of composite spectral samples built directly
from the data, with accurate $(G-\cal{R})$ colours and $\cal{R}$
magnitudes, do not reproduce the $(U_n-G)$ vs. $(G-\cal{R})$
distribution of the data when assuming zero LyC flux.  A large
discrepancy between the colour distribution of the composite spectral
samples and the data remains after accounting for potential $U_n$
filter redleak, `$U_n$-drop' magnitude assignments,
wavelength-dependent flux loss in the spectroscopic observations, and
lower redshift line-of-sight contamination.  The discrepancy can be
resolved by an increase in $U_n$-band flux from the galaxies which, at
$z\sim$ 3--4, corresponds to restframe LyC flux.

\item With the addition of LyC flux to the composite spectra, the
evolutionary tracks accurately reproduce the LyC flux detections in
the literature.  Using only the colours and \lya\ EWs of galaxies, the
composite spectra can accurately predict their redshifts, properties,
and LyC flux.

\item We define \robsl, the fraction of observed LyC (from restframe
wavelength $\lambda$ to 912\AA) to non-ionising UV flux, as a more
intuitive means to describe the observed LyC flux in galaxy
spectroscopy and photometry.  \robsu\ denotes the ratio of LyC
integrated from the shortest wavelength probed by the $U_n$ filter for
the redshift of the data to 912\AA.

\item We compare the \lya\ EWs and colours of the 794 spectra of
\citet{shapley03} to the composite spectra evolutionary tracks and
measure their expected LyC flux as a statistical sample of the \zzz\
LBG population.  We find an average ${\mbox{\robsu}=5.0^{+1.0}_{-0.4}~
(4.1^{+0.5}_{-0.3})}$ percent and a corresponding
${\mbox{\fesc}=10.5^{+2.0}_{-0.8}~(8.6^{+1.0}_{-0.6})}$ percent for
LBGs at ${z\sim3}$ meeting the standard colour-selection criteria
(contamination corrected).  To derive \fesc\ from \robsu, we assume a
$z=$ 3 IGM attenuation factor of 2.4, an intrinsic UV continuum to LyC
luminosity density fraction of 3.0, and the extinction law of
\citet{calzetti00} and the $E(B-V)$ values of the composite spectra
quartiles.

\item The LyC measurements of the individual spectra find that LBGs
with stronger \lya\ EW have higher LyC flux.  The cause of this
effect, in part, is the result of the restriction of objects to the
standard colour-selection region.  However, to match the photometric
distribution of the data we are forced to assign a higher level of LyC
flux to composite spectral samples with higher \lya\ EW which suggests
that part of this relationship may also be intrinsic.  Both \lya\ and
LyC radiation transfer are sensitive to H\textsc{i}.  Thus, it may be
that galaxies with \lya\ in emission, and their corresponding
properties such as bluer continua, weaker ISM metal-line features,
compact morphology, field-like environments, and potentially higher
merger rates, or their orientation cause higher LyC escape fractions.

\item Both the distribution of the data and the evolutionary tracks of
the modified composite spectra require a significant fraction of
$z\sim$ 3--4 galaxies residing outside the standard colour-selection
region.  Galaxies at $z\sim$ 3--4 with colours outside the standard
LBG selection region have been spectroscopically confirmed by the VVDS
magnitude-limited survey and reside in the locations predicted by the
evolutionary tracks of the modified composite spectra.

\item Many of the galaxies found outside the standard colour-selection
region are missed by other (i.e., lower redshift) selection criteria.
The `missed' galaxies are expected to have moderate to high levels of
escaping LyC flux from their bluer $(U_n-G)$ colours and from the
predictions of the composite spectra evolutionary tracks and are
termed Lyman continuum galaxies (LCGs).  LCGs are estimated to
increase the census of \zzz\ galaxies by a factor of $\sim$1.5.

\item Matching the composite spectral samples to the photometric data
distribution, we estimate \robsu\ = 15.8 $\pm3.3$ and \fesc\ = 33.0
$\pm6.9$ percent for LCGs and \robsu\ = 7.6 $\pm1.7$ and \fesc\ = 15.9
$\pm3.5$ percent for the combined \zzz\ LBG + LCG population using the
same assumptions for LBGs above.

\item LCGs reside in a region of colour-colour space with a high
density of low redshift sources.  We find that examining the colours
of \zzzz\ LCGs on the $(U_n-G)$ vs. $(G-\cal{R})$ plane, typically
used for \zzz\ colour selection, provides one means to increase the
selection efficiency.  A parallel analysis using deep near-UV imaging
can be used to efficiently select \zzz\ galaxies.  

\end{enumerate}

Our analysis has helped confirm that conventional LBG colour-selection
criteria do not include the entire \zzz\ galaxy population.  This
point has been know since the criteria conception and has been
reinforced by spectroscopic identifications of $z\sim$ 3--4 galaxies
outside the standard \zzz\ selection region.  The modified composite
spectra provide a physical basis for the existence of LCGs and
indicate that they require flux in the bluest filter (i.e., the
`drop-out' filter) and, thus, have high levels of LyC flux.

Efficient selection of $z\sim$ 3--4 remains a challenge, however, deep
broadband and medium band infrared (restframe optical) imaging
provides a promising approach.  The number density of LBGs remains to
be quantified as well as properties that may facilitate high escaping
flux, including their UV continua slopes, SEDs, line-of-sight \lya\
forest decrements, average outflow strengths, orientation, and the
effects of environment and interactions.  The measured LyC flux for
LCGs, when added to the LyC flux measured here for LBGs, will help
clarify the full contribution of galaxies to the reionisation of the
Universe.


\section*{Acknowledgments}

The authors would like to thank C. C. Steidel for helpful discussions.
JC acknowledges the support of the Australian Research Council Future
Fellowship grant FT 130101219.  ERW acknowledges the support of the
Australian Research Council grant DP 1095600 and CGD acknowledges the
support by the Victorian Government.  This research uses data from the
VIMOS VLT Deep Survey, obtained from the VVDS database operated by
Cesam, Laboratoire d'Astrophysique de Marseille, France.  This work
has made use of the VizieR catalogue access tool, CDS, Strasbourg,
France.  Photometry presented here are based on observations obtained
with MegaPrime/MegaCam, a joint project of CFHT and CEA/IRFU, at the
Canada-France-Hawaii Telescope (CFHT) which is operated by the
National Research Council (NRC) of Canada, the Institut National des
Science de l'Univers of the Centre National de la Recherche
Scientifique (CNRS) of France, and the University of Hawaii. This work
is based in part on data products produced at Terapix available at the
Canadian Astronomy Data Centre as part of the Canada-France-Hawaii
Telescope Legacy Survey, a collaborative project of NRC and CNRS.


\label{lastpage}


\begin{thebibliography}{}

\bibitem[Barkana \& Loeb(1999)]{barkana99} Barkana, R. \& Loeb, A.\
  1999, \apj, 523, 54
\bibitem[Becker, Rauch, \& Sargent(2007)]{becker07} Becker, G.~D.,
  Rauch, M., \& Sargent, W.~L.~W.\ 2007, \apj, 662, 72
\bibitem[Becker et al.(2013)]{becker13} Becker, G.~D., Hewett, P.~C.,
  Worseck, G., \& Prochaska, J.~X.\ 2013, \mnras, 430, 2067
\bibitem[Bertin \& Arnouts(1996)]{ba96} Bertin, E. \& Arnouts,
  S. 1996, A\&AS, 117, 393
\bibitem[Bielby et al.(2011)]{bielby11} Bielby, R.~M., Shanks, T.,
  Weilbacher, P.~M., et al.\ 2011, \mnras, 414, 2
\bibitem[Bielby et al.(2013)]{bielby13} Bielby, R., Hill, M.~D.,
  Shanks, T., et al.\ 2013, \mnras, 430, 425
\bibitem[Bland-Hawthorn \& Maloney(1999)]{blandhawthorn99}
  Bland-Hawthorn, J. \& Maloney, P.R. 1999, ApJ, 510, L33
\bibitem[Bouwens et al.(2007)]{bouwens07} Bouwens, R.~J., Illingworth,
  G.~D., Franx, M., \& Ford, H.\ 2007, \apj, 670, 928
\bibitem[Bouwens et al.(2011)]{bouwens11} Bouwens, R.~J., Illingworth,
  G.~D., Oesch, P.~A., et al.\ 2011, \apj, 737, 90
\bibitem[Bruzual \& Charlot(2003)]{bruzual03} Bruzual, G., \& Charlot,
  S.\ 2003, \mnras, 344, 1000
\bibitem[Bullock, Kravtsov, \& Weinberg(2000)]{bullock00} Bullock,
  J.~S. and Kravtsov, A.~V. \& Weinberg, D.~H.\ 2000, \apj, 539, 517
\bibitem[Calverley et al.(2011)]{calverley11} Calverley, A.~P.,
  Becker, G.~D., Haehnelt, M.~G., \& Bolton, J.~S.\ 2011, MNRAS, 412,
  2543
\bibitem[Calzetti et al.(2000)]{calzetti00} Calzetti, D., Armus, L.,
  Bohlin, R.~C., et al.\ 2000, \apj, 533, 682
\bibitem[Chapman et al.(2005)]{chapman05} Chapman, S.~C., Blain,
  A.~W., Smail, I., \& Ivison, R.~J.\ 2005, \apj, 622, 772
\bibitem[Cooke et al.(2005)]{cooke05} Cooke, J., Wolfe, A.~M.,
  Prochaska, J.~X., \& Gawiser, E.\ 2005, \apj, 621, 596
\bibitem[Cooke et al.(2006)]{cooke06} Cooke, J., Wolfe, A.~M.,
  Gawiser, E., \& Prochaska, J.~X.\ 2006, \apj, 652, 994
\bibitem[Cooke(2009)]{cooke09} Cooke, J.\ 2009, ApJL, 704, L62
\bibitem[Cooke et al.(2010)]{cooke10} Cooke, J., Berrier, J. C.,
  Barton, E. J., Bullock, J. S., \& Wolfe, A. M. ~2010, MNRAS, 403,
  1020
\bibitem[Cooke, Omori, \& Ryan-Weber(2013)]{cooke13} Cooke, J., Omori,
  Y, \& Ryan-Weber, E. V.\ 2013, MNRAS, 433, 2122
\bibitem[Daddi et al.(2004)]{daddi04} Daddi, E., Cimatti, A., Renzini,
  A., et al.\ 2004, \apj, 617, 746
\bibitem[Dove \& Shull(1994)]{dove94} Dove, J.B. \& Shull, J.M.,
  1994, ApJ, 430, 222
\bibitem[Efstathiou(1992)]{efstathiou92} Efstathiou, G.\ 1992, MNRAS,
  256, 43
\bibitem[Ellis et al.(2013)]{ellis13} Ellis, R.~S., McLure, R.~J.,
  Dunlop, J.~S., et al.\ 2013, \apjl, 763, L7
\bibitem[Fan et al.(2006)]{fan06} Fan, X., Strauss, M.~A., Becker,
  R.~H., White, R.~L., Gunn, J.~E., Knapp, G.~R., Richards, G.~T.,
  Schneider, D.~P., Brinkmann, J., \& Fukugita, M.\ 2006, \aj, 132,
  117
\bibitem[Ferguson et al.(2004)]{ferguson04} Ferguson, H.~C.,
  Dickinson, M., Giavalisco, M., et al.\ 2004, \apjl, 600, L107
\bibitem[Fernandez \& Shull(2011)]{fernandez11} Fernandez, E.~R. \&
  Shull, J.~M.\ 2011, \apj, 731, 20
\bibitem[Finkelstein et al.(2010)]{finkelstein10} Finkelstein, S.~L.,
  Papovich, C., Giavalisco, M., et al.\ 2010, \apj, 719, 1250
\bibitem[Finlator et al.(2012)]{finlator12} Finlator, K., Oh, S.~P.,
  {\"O}zel, F., \& Dav{\'e}, R.\ 2012, \mnras, 427, 2464
\bibitem[Fontanot, Cristiani, \& Vanzella(2012)]{fontanot12} Fontanot,
  F., Cristiani, S., \& Vanzella, E.\ 2012, MNRAS, 425, 1413
\bibitem[Fukugita et al.(1996)]{fukugita96} Fukugita, M., Ichikawa,
  T., Gunn, J. E., Doi, M., Shimasaku, K., \& Schneider, D. P. 1996,
  AJ, 111, 1748
\bibitem[Giallongo \& Cristiani(1990)]{giallongo90} Giallongo, E., \&
  Cristiani, S.\ 1990, \mnras, 247, 696
\bibitem[Gnedin et al.(2008)]{gnedin08} Gnedin, N.~Y., Kravtsov,
  A.~V., \& Chen, H.-W.\ 2008, \apj, 672, 765
\bibitem[Guhathakurta et al.(1990)]{guhathakurta90} Guhathakurta, P.,
  Tyson, J.~A., \& Majewski, S.~R.\ 1990, \apjl, 357, L9
\bibitem[Haardt \& Madau(2012)]{haardt12} Haardt, F. \& Madau, P.\
  2012, \apj, 746, 125
\bibitem[Hopkins, Richards, \& Hernquist(2007)]{hopkins07} Hopkins,
  P.~F., Richards, G.~T., Hernquist, L.\ 2007, \apj, 654, 731
\bibitem[Iliev et al.(2007)]{iliev07} Iliev, I.~T., Mellema, G.,
  Shapiro, P.~R., \& Pen, U.-L.\ 2007, MNRAS, 376, 534
\bibitem[Inoue et al.(2005)]{inoue05} Inoue, A.~K., Iwata, I.,
  Deharveng, J.-M., Buat, V., \& Burgarella, D.\ 2005, \aap, 435, 471
\bibitem[Iwata et al.(2009)]{iwata09} Iwata, I., Inoue, A.~K.,
  Matsuda, Y., et al.\ 2009, \apj, 692, 1287
\bibitem[Jiang et al.(2008)]{jiang08} Jiang, L., Fan, X., Annis, J.,
  Becker, R.~H., White, R.~L., Chiu, K., Lin, H., Lupton, R.~H.,
  Richards, G.~T., Strauss, M.~A., Jester, S., \& Schneider, D.~P.\
  2008, \aj, 135, 1057
\bibitem[Jones et al.(2012)]{jones12} Jones, T., Stark, D.~P., \&
  Ellis, R.~S.\ 2012, \apj, 751, 51
\bibitem[Komatsu et al.(2011)]{komatsu11} Komatsu, E., Smith, K.~M.,
  Dunkley, J., Bennett, C.~L., Gold, B., Hinshaw, G., Jarosik, N.,
  Larson, D., Nolta, M.~R., Page, L., Spergel, D.~N., Halpern, M.,
  Hill, R.~S., Kogut, A., Limon, M., Meyer, S.~S., Odegard, N.,
  Tucker, G.~S., Weiland, J.~L., Wollack, E., \& Wright, E.~L.\ 2011,
  \apjs, 192, 18
\bibitem[Kuhlen \& Faucher-Gigu{\`e}re(2012)]{kuhlen12} Kuhlen, M., \&
  Faucher-Gigu{\`e}re, C.-A.\ 2012, \mnras, 423, 862
\bibitem[Law et al.(2007)]{law07} Law, D.~R., Steidel, C.~C., Erb,
  D.~K., et al.\ 2007, \apj, 656, 1
\bibitem[Law et al.(2012)]{law12} Law, D.~R., Steidel, C.~C., Shapley,
  A.~E., et al.\ 2012, \apj, 759, 29
\bibitem[Le F{\`e}vre et al.(2005a)]{lefevre05a} Le F{\`e}vre, O.,
  Vettolani, G., Garilli, B., et al.\ 2005a, \aap, 439, 845
\bibitem[Le F{\`e}vre et al.(2005b)]{lefevre05b} Le F{\`e}vre, O., et
  al.\ 2005b, Nature, 437, 519
\bibitem[Le F{\`e}vre et al.(2013a)]{lefevre13a} Le Fevre, O., Cassata,
  P., Cucciati, O., et al.\ 2013a, \aap, 559, A14
\bibitem[Le F{\`e}vre et al.(2013b)]{lefevre13b} Le Fevre, O., Cassata,
  P., Cucciati, O., et al.\ 2013b, arXiv:1307.6518
\bibitem[Lu \& Zuo(1994)]{lu94} Lu, L., \& Zuo, L.\ 1994, \apj, 426,
  502
\bibitem[Ly et al.(2011)]{ly11} Ly, C., Malkan, M.~A., Hayashi, M., et
  al.\ 2011, \apj, 735, 91
\bibitem[Madau, Haardt, \& Rees(1999)]{madau99} Madau, P., Haardt,
  F. \& Rees, M.~J.\ 1999, \apj, 514, 648
\bibitem[McQuinn, Oh, \& Faucher-Gigu{\`e}re(2011)]{mcquinn11}
  McQuinn, M., Oh, S.~P., \& Faucher-Gigu{\`e}re, C.-A.\ 2011, \apj,
  743, 82
\bibitem[Nestor et al.(2013)]{nestor13} Nestor, D.~B., Shapley, A.~E.,
  Kornei, K.~A., Steidel, C.~C., \& Siana, B.\ 2013, \apj, 765, 47
\bibitem[Oke et al.(1995)]{oke95} Oke, J. B., Cohen, J. G., Carr, M.,
  Cromer, J., Dingizian, A., Harris, F. H., Labrecque, S., Lucinio,
  R., Schaal, W., Epps, H., \& Miller, J.  1995, PASP, 107, 375
\bibitem[Paltani et al.(2007)]{paltani07} Paltani, S., Le F{\`e}vre,
  O., Ilbert, O., et al.\ 2007, \aap, 463, 873
\bibitem[Pawlik, Schaye, \& van Scherpenzeel(2009)]{pawlik09} Pawlik,
  A.~H., Schaye, J., \& van Scherpenzeel, E.\ 2009, MNRAS, 394, 1812
\bibitem[Prochaska et al.(2009)]{prochaska09} Prochaska, J.~X.,
  Worseck, G., \& O'Meara, J.~M.\ 2009, \apjl, 705, L113
\bibitem[Putman et al.(2003)]{putman03} Putman, M.E. and Bland-Hawthorn,
  J. and Veilleux, S. and Gibson, B.K. and Freeman, K.C. and Maloney,
  P.R., ApJ, 597, 948
\bibitem[Razoumov \& Sommer-Larsen(2006)]{razoumov06} Razoumov, A.~O.,
  \& Sommer-Larsen, J.\ 2006, \apjl, 651, L89
\bibitem[Razoumov \& Sommer-Larsen(2010)]{razoumov10} Razoumov, A.~O.,
  \& Sommer-Larsen, J.\ 2010, \apj, 710, 1239
\bibitem[Reichart(2001)]{reichart01} Reichart, D.~E.\ 2001, \apj, 553,
  235
\bibitem[Ricotti \& Shull(2000)]{ricotti00} Ricotti, M. \& Shull,
  J.~M.\ 2000, \apj, 542, 548
\bibitem[Robertson et al.(2013)]{robertson13} Robertson, B.~E.,
  Furlanetto, S.~R., Schneider, E., Charlot, S., Ellis, R.~S., Stark,
  D.~P., McLure, R.~J., Dunlop, J.~S., Koekemoer, A., Schenker, M.~A.,
  Ouchi, M., Ono, Y., Curtis-Lake, E., Rogers, A.~B., Bowler,
  R.~A.~A. \& Cirasuolo, M.\ 2013, \apj, 768, 71
\bibitem[Ryan-Weber et al.(2009)]{ryan-weber09} Ryan-Weber, E.~V.,
  Pettini, M., Madau, P., \& Zych, B.~J.\ 2009, \mnras, 395, 1476
\bibitem[Shapley et al.(2003)]{shapley03} Shapley, A. E.,
  Steidel,  C. C., Adelberger, K. L., \& Pettini, M. 2003, ApJ, 588, 65
\bibitem[Shapley et al.(2006)]{shapley06} Shapley, A. E., Steidel,
  C. C., Pettini, M., \& Adelberger, K. L.\ 2006, ApJ, 651, 688
\bibitem[Shull et al.(2012)]{shull12} Shull, J.~M., Harness, A.,
  Trenti, M., \& Smith, B.~D.\ 2012, \apj, 747, 100
\bibitem[Somerville(2002)]{somerville02} Somerville, R.~S.\ 2002,
  \apjl, 572, 23
\bibitem[Songaila et al.(1990)]{songaila90} Songaila, A., Cowie,
  L.~L., \& Lilly, S.~J.\ 1990, \apj, 348, 371
\bibitem[Spitler et al.(2014)]{spitler14} Spitler, L., et al.\ 2014,
  ApJL, {\it submitted}, arXiv:
\bibitem[Steidel \& Hamilton(1992)]{steidel92} Steidel, C.~C., \&
  Hamilton, D.\ 1992, \aj, 104, 941
\bibitem[Steidel \& Hamilton(1993)]{steidel93} Steidel, C.~C., \&
  Hamilton, D.\ 1993, \aj, 105, 2017
\bibitem[Steidel et al.(1996)]{steidel96} Steidel, C. C., Giavalisco,
  M., Pettini, M., Dickinson, M., \& Adelberger, K. L. 1996, ApJ, 462,
  17
\bibitem[Steidel et al.(1999)]{steidel99} Steidel, C.~C., Adelberger,
  K.~L., Giavalisco, M., Dickinson, M., \& Pettini, M.\ 1999, \apj,
  519, 1
\bibitem[Steidel et al.(2001)]{steidel01} Steidel, C. C., Pettini, M.,
  \& Adelberger, K. L.\ 2001, ApJ, 546, 665
\bibitem[Steidel et al.(2003)]{steidel03} Steidel, C. C., Adelberger,
  K. L., Shapley, A. E., Pettini, M., Dickinson, M., \& Giavalisco,
  M. 2003, ApJ, 592, 728
\bibitem[Steidel et al.(2004)]{steidel04} Steidel, C. C., Shapley,
  A. E., Pettini, M., Adelberger, K. L., Erb, D. K., Reddy, N. A., \&
  Hunt, M. P.\ 2004, ApJ, 604, 534
\bibitem[Trenti et al.(2010)]{trenti10} Trenti, M., Stiavelli, M.,
  Bouwens, R.~J., et al.\ 2010, \apjl, 714, L202
\bibitem[van Dokkum et al.(2006)]{vandokkum06} van Dokkum, P.~G.,
  Quadri, R., Marchesini, D., et al.\ 2006, \apjl, 638, L59
\bibitem[Wood \& Loeb(2000)]{wood00} Wood, K. \& Loeb, A.~ 2000, \apj,
  545, 86
\bibitem[Wyithe \& Bolton(2011)]{wyithe11} Wyithe, J.~S.~B. \&
  Bolton, J.~S.\ 2011, MNRAS, 412, 1926
\bibitem[Yajima et al.(2011)]{yajima11} Yajima, H., Choi, J.-H., \&
  Nagamine, K.\ 2011, \mnras, 412, 411
\bibitem[Zahn et al.(2012)]{zahn12} Zahn, O., Reichardt, C.~L., Shaw,
  L., et al.\ 2012, \apj, 756, 65

\end{thebibliography}
\end{document}